\documentclass[reprint,10pt,twocolumn,pra,superscriptaddress,showpacs,aps]{revtex4-1}   	

\usepackage{graphicx}
\usepackage{amsmath}
\usepackage{amssymb}

\usepackage[normalem]{ulem} 
\usepackage[usenames,dvipsnames]{color}

\usepackage[free-standing-units=true,use-xspace=true,retain-unity-mantissa = false]{siunitx}
\DeclareSIUnit\gauss{G}

\newcommand{\G}{\gauss}

\newcommand{\eV}{\electronvolt}
\newcommand{\mbar}{\milli\bar}
\newcommand{\K}{\kelvin}
\newcommand{\mK}{\milli\kelvin}
\newcommand{\uK}{\micro\kelvin}
\newcommand{\sr}{\steradian}
\newcommand{\m}{\meter}
\newcommand{\km}{\kilo\meter}
\newcommand{\mm}{\milli\meter}
\newcommand{\cm}{\centi\meter}
\newcommand{\um}{\micro\meter}
\newcommand{\nm}{\nano\meter}
\newcommand{\mrad}{\milli\radian}

\newcommand{\s}{\second}
\newcommand{\ms}{\milli\second}
\newcommand{\us}{\micro\secind}
\newcommand{\ns}{\nano\second}
\newcommand{\ps}{\pico\s}

\newcommand{\Hz}{\hertz}

\newcommand{\MHz}{\mega\hertz}

\newcommand{\kV}{\kilo\volt}

\newcommand{\W}{\watt}

\DeclareSIUnit\atoms{atoms}

\begin{document}
\title{Bose-Einstein condensate of metastable helium for quantum correlation experiments}
\author{Michael \surname{Keller}}
\email[Corresponding author:\ ]{michael.keller@univie.ac.at}
\affiliation{Institute for Quantum Optics and Quantum Information, Austrian Academy of Sciences, Boltzmanngasse 3, 1090 Vienna, Austria}
\affiliation{Faculty of Physics, University of Vienna, Boltzmanngasse 5, 1090 Vienna, Austria}
\author{Mateusz \surname{Kotyrba}}
\affiliation{Institute for Quantum Optics and Quantum Information, Austrian Academy of Sciences, Boltzmanngasse 3, 1090 Vienna, Austria}
\affiliation{Faculty of Physics, University of Vienna, Boltzmanngasse 5, 1090 Vienna, Austria}
\author{Florian \surname{Leupold}}
\altaffiliation[Present address:\ ]{Institute for Quantum Electronics, ETH Z{\"u}rich, Otto-Stern-Weg 1, 8093 Z{\"u}rich, Switzerland}
\affiliation{Institute for Quantum Optics and Quantum Information, Austrian Academy of Sciences, Boltzmanngasse 3, 1090 Vienna, Austria}
\author{Mandip \surname{Singh}}
\altaffiliation[Present address:\ ]{Department of Physical Sciences, Indian Institute of Science Education and Research Mohali, India}
\affiliation{Institute for Quantum Optics and Quantum Information, Austrian Academy of Sciences, Boltzmanngasse 3, 1090 Vienna, Austria}
\author{Maximilian \surname{Ebner}}
\affiliation{Institute for Quantum Optics and Quantum Information, Austrian Academy of Sciences, Boltzmanngasse 3, 1090 Vienna, Austria}
\affiliation{Faculty of Physics, University of Vienna, Boltzmanngasse 5, 1090 Vienna, Austria}
\author{Anton \surname{Zeilinger}}
\affiliation{Institute for Quantum Optics and Quantum Information, Austrian Academy of Sciences, Boltzmanngasse 3, 1090 Vienna, Austria}
\affiliation{Faculty of Physics, University of Vienna, Boltzmanngasse 5, 1090 Vienna, Austria}


\begin{abstract}
We report on the realization of Bose-Einstein condensation of metastable helium-4. After exciting helium to its metastable state in a novel pulse-tube cryostat source, the atomic beam is collimated and slowed. We then trap several $\num{1e8}$ atoms in a magneto-optical trap. For subsequent evaporative cooling, the atoms are transferred into a magnetic trap. Degeneracy is achieved with typically a few $\num{1e6}$ atoms. For detection of atomic correlations with high resolution, an ultrafast delay-line detector has been installed. Consisting of four quadrants with independent readout electronics that allow for true simultaneous detection of atoms, the detector is especially suited for quantum correlation experiments that require the detection of correlated subsystems. We expect our setup to allow for the direct demonstration of momentum entanglement in a scenario equivalent to the Einstein-Podolsky-Rosen gedanken experiment. This will pave the way to matter-wave experiments exploiting the peculiarities of quantum correlations.
\end{abstract}
\maketitle

\section{Introduction}

Since the first realization of a Bose-Einstein condensate (BEC)~\cite{Anderson:1995,Davis:1995,Bradley:1995} the field of dilute atomic gases has grown substantially and has opened up the possibility of novel research in many areas of physics. Among numerous developments, it enabled studies of the transition from the superfluid to the Mott insulator phase~\cite{Greiner:2002}, the BEC-BCS crossover~\cite{Jochim:2003,Zwierlein:2003,Greiner:2003}, and quantum atom optics~\cite{Mewes:1997,Deng:1999,Perrin:2007}. While most of these experiments were limited to the observation of collective properties of the ensemble, more recent detection techniques allow for the resolution of single sites in optical lattices~\cite{Gericke:2008,Bakr:2009,Sherson:2010} or even the detection of individual atoms from an atomic cloud~\cite{Schellekens:2005,Buecker:2009}. The first method of this kind exploited the high internal energy of metastable states in noble gas atoms (several $\si{\eV}$ compared to about $\SI{1e-10}{\eV}$ of kinetic energy in a BEC) to efficiently detect single atoms. In particular, BECs of metastable helium (He*)~\cite{Robert:2001,Santos:2001,Vassen:2012} have extended the field of atomic physics experiments to the single-particle level~\cite{Schellekens:2005,Jeltes:2007,Hodgman:2011}.

Single-atom detection opens the path to performing matter-wave analogs~\cite{Ferris:2008,Ferris:2009,Kofler:2012} of some of the pioneering photon-entanglement experiments~\cite{Rarity:1990,Ou:1990,Ou:1992}. Where spontaneous parametric down-conversion (SPDC) commonly serves as a source for entangled photons, four-wave mixing (FWM) can create entanglement in the atomic case. FWM has been realized within BECs~\cite{Deng:1999,Vogels:2002,Vogels:2003} and, in combination with single-atom detection, revealed correlations outperforming the classical ones~\cite{Perrin:2007,Kheruntsyan:2012}.

It is expected that the FWM process leads to momentum entangled particles which would be a three-dimensional analog of the Einstein-Podolsky-Rosen (EPR) gedanken experiment~\cite{Einstein:1935}. The EPR argument involves entanglement of freely moving, massive particles in their external degrees of freedom, and its realization is a long-standing quest of experimental quantum physics. Not only will its realization be a useful tool for a direct comparison with the photon entanglement experiments, but can potentially boost the sensitivity of matter-wave interferometers. From there on, specific features of matter, such as the access to Fermi statistics and its strong susceptibility to gravity, will extend the playground of entanglement beyond the possibilities of photonics. 

\begin{figure*}[t]
\centering
\includegraphics[width=\textwidth]{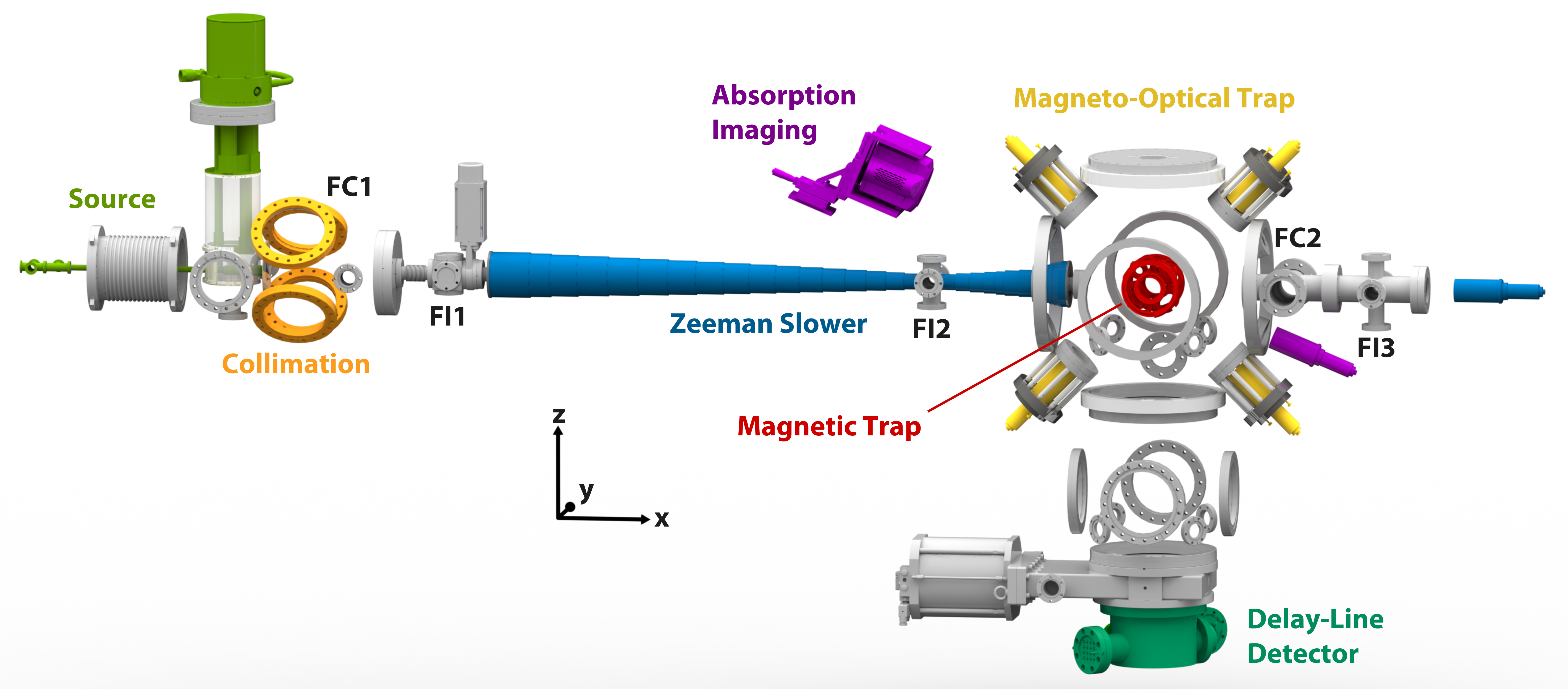}
\caption{(Color online) Overview of the vacuum system (starting from the left): The atoms are initially excited in the source (helium inlet tube and closed-cycle cryostat marked in green). The atoms get collimated by laser beams that enter through the windows drawn in orange and are decelerated in the Zeeman slower (highlighted in blue, as well as the telescope for the slowing laser beam). Four of the telescopes for the magneto-optical trap lasers are displayed in yellow (the two along the y-direction are not shown). Bose-Einstein condensation is achieved within the magnetic trap (coils shown in red) after forced evaporative cooling by radio frequency. The atoms can be imaged by absorption imaging (EMCCD camera and telescope in purple) or impact onto the single-particle resolving delay-line detector (located in the vacuum vessel in dark green). The figure is drawn to scale but some components (e.g., vacuum chambers and vacuum pumps) are not displayed. \emph{FC} marks the positions where the two installed Faraday cup detectors can be inserted into the atomic beam path, the possible access points for fluorescence imaging of the atomic beam are marked with \emph{FI}.}
\label{fig:overview}
\end{figure*}

Here we report on the realization of a BEC with typically several $10^{6}$ He* atoms. Our experimental setup (see Fig.~\ref{fig:overview}) combines optical detection and a delay-line detector. The latter is crucial for single-atom detection and eventual quantum correlation measurements. To achieve this, it is designed for high count rate, good resolution, and the possibility of true simultaneous detection with its four independent quadrants, a novel feature that will be especially useful for experiments in schemes with number squeezed states and a highly directional output as, for example, described in~\cite{Dall:2009,Buecker:2011}. The structure of this paper follows the path of the atoms -- from source to detection. First, a bright, continuous beam of metastable helium atoms is produced, precooled using a closed-cycle cryostat, and collimated by means of laser cooling (Sec.~\ref{s:source}). Next, longitudinal slowing is achieved in a Zeeman slower (Sec.~\ref{s:slower}), which makes magneto-optical trapping of the atoms possible (Sec.~\ref{s:mot}). The atomic cloud is then transferred into a magnetic trap (Sec.~\ref{s:mt}), where Bose-Einstein condensation is reached after evaporative cooling (Sec.~\ref{s:bec}). Finally, we present our detection schemes, especially the delay-line detector (DLD) for single-atom detection (Sec.~\ref{s:detection}).

\section{Source}\label{s:source}

The very first part of any metastable helium experiment is a source where helium atoms are transferred to the metastable $2^{3}\text{S}_{1}$ state. Due to selection rules there are no allowed electric dipole and quadrupole transitions between this state and the ground state \cite{Drake:1971}. This makes the highly energetic state ($\SI{19.8}{\eV}$ above the ground state) hard to excite but also very long lived, that is, metastable. The radiative lifetime of almost $\SI{8000}{s}$ \cite{Moos:1973,Hodgman:2009} allows us to consider this state as the effective ground state, with two laser-accessible closed transitions. These transitions allow for optical manipulation of He* atoms.

\begin{figure*}
\centering
\includegraphics[width=0.7\textwidth]{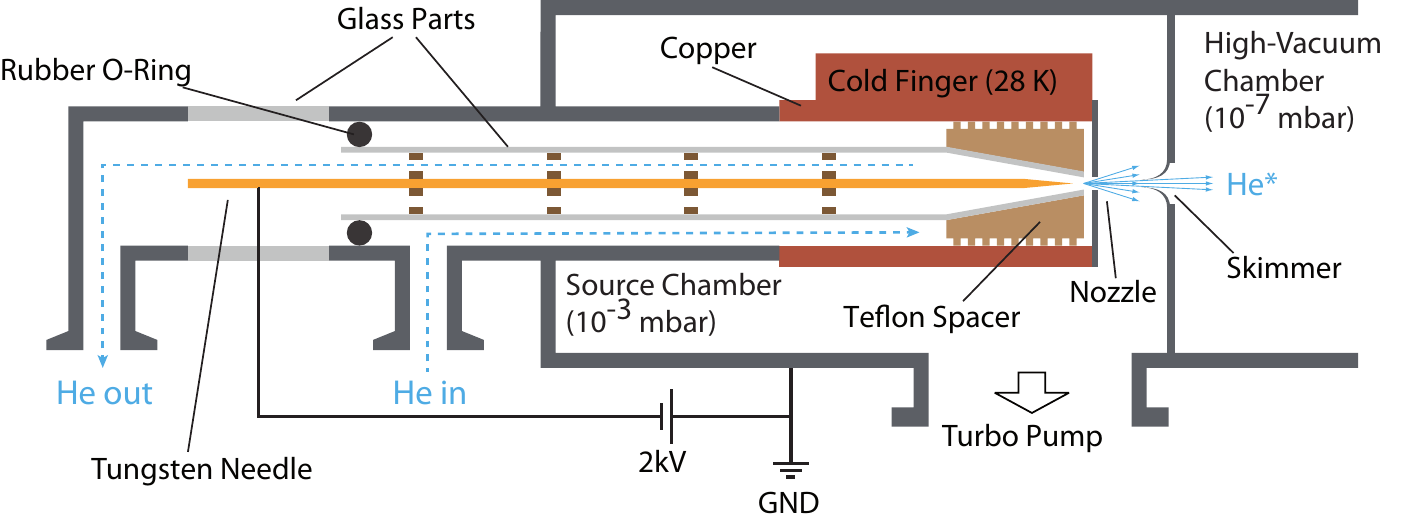}
\caption{(Color online) Schematic of the metastable helium source. Helium enters the vacuum system and flows to the front of the vacuum tube where a helically grooved teflon spacer ensures good contact with a copper tube cooled by a cryostat. A part of the atoms is excited in an electric discharge between the tungsten needle and nozzle. While a fraction of the atoms enters the high-vacuum chamber after supersonic expansion through the nozzle and spatial filtering by a skimmer, the rest gets pumped out at the back of the source. Ceramic spacers align the tungsten needle in the glass tube; glass vacuum parts prevent parasitic discharges.}
\label{fig:source}
\end{figure*}

We have built a supersonic, cold-cathode, dc discharge source that yields high flux and continuous output. As in the design of Kawanaka et al.~\cite{Kawanaka:1993}, the metastables are produced by electron impact in a reverse flow scheme, in which the gas flow at the discharge position is opposite to the output direction of the source, which helps to reduce the velocity of the emitted atoms (Fig.~\ref{fig:source}). Helium enters the vacuum system through a needle valve ($\SI{3}{\mbar}$ inlet pressure). The gas flows between a glass tube and a steel tube to the cold finger. The cold finger is held at a temperature of approximately $\SI{28}{\K}$ to reduce the kinetic energy of the helium by contact cooling. The cold atoms then pass a strong electric field between a tungsten needle kept at $\SI{-2}{\kV}$ and a grounded nozzle, where they are ionized by field ionization and a plasma is created. While some of the gas expands supersonically through a $\SI{0.5}{\mm}$-wide nozzle, the rest of the gas flows back through the glass tube and is pumped out of the system. By recombination of the helium ions with free electrons, neutral helium atoms in electronically excited states are created. Through various cascades, a fraction of the atoms ends up in the energetically lowest triplet state of helium $2^{3}\text{S}_{1}$. Together with photons, ions, and helium atoms in other internal states, a fraction of these metastables enters the high vacuum chamber ($\SI{1e-7}{\mbar}$ when operating) through a conically shaped skimmer with a $\SI{0.5}{\mm}$-wide opening.

The main design parameters of a He* source for cold atom experiments are flux and velocity. While the former should be as high as possible, the latter should be as low as possible. The low velocity ensures a short distance for the subsequent deceleration and thereby increases the capture efficiency of the magneto-optical trap (MOT). Since in discharge sources a higher flux is mainly achieved by intensifying the plasma (either by increasing the helium inflow or the discharge voltage), this also causes more heating and thereby higher velocities of the atoms. Some of the He* sources reported in the literature are run at room temperature while others are cooled with liquid nitrogen or even liquid helium. Our design uses a closed-cycle two-stage pulse-tube cooler that cools our source to temperatures significantly below liquid-nitrogen temperatures ($<\SI{30}{\K}$ during operation) while still being able to deliver a significant cooling power of about $\SI{2}{\W}$ at those temperatures. With this, we can tune the discharge parameters to generate an atomic beam with a flux in the range of \SIrange[range-units=brackets,range-phrase = --,fixed-exponent=14,scientific-notation=fixed,retain-unity-mantissa = true]{1e14}{8e14}{\atoms\per\sr\per\s} and velocities between $\SI{650}{\m/\s}$ and $\SI{1000}{\m/\s}$. There are three main contributions to the mean velocity of the atomic beam that can be achieved: firstly, the temperature of the cryostat and the available cooling power, secondly, the heating by the plasma and finally, an increase in velocity due to the supersonic expansion after the skimmer. The first two aspects correlate the mean velocity with the achievable flux since the temperature of a given cryostat increases with the gas load and a more intense plasma further increases the velocity. We usually operate our source at $\SI{4e14}{\atoms\per\sr\per\s}$ and a peak velocity of $\SI{800}{\m/\s}$ by applying a discharge voltage of $\SI{-2}{\kV}$ and adjusting the helium inlet to around $\SI{3}{mbar}$. The beam intensities were measured with a Faraday cup detector in which electrons are ejected upon collision with a stainless steel plate by the high internal energy of the He* atoms. In order to determine the velocities, a time-of-flight technique was used.

\begin{figure}
\centering
\includegraphics[width=\columnwidth]{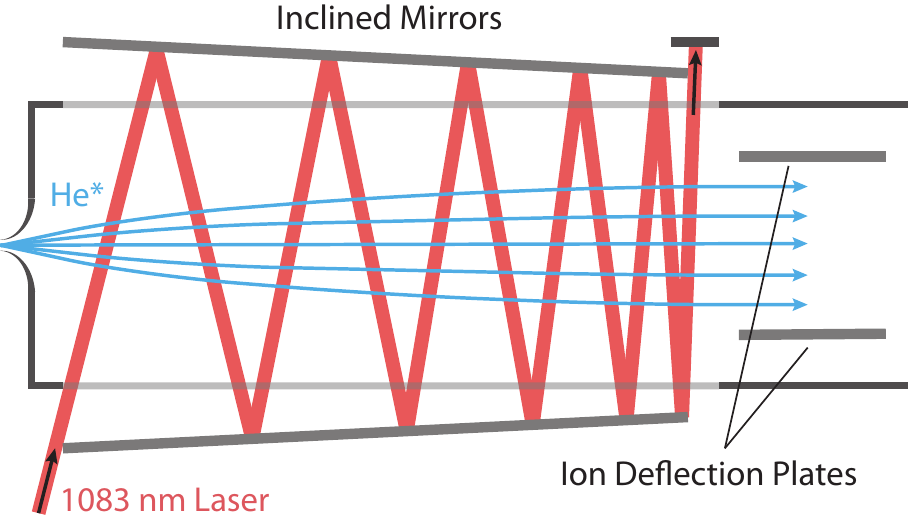}
\caption{(Color online) Scheme of the laser beam alignment between one pair of the collimation mirrors. A $\SI{6}{\mm}$-wide laser beam is reflected 12 times between two mirrors that are inclined at a small angle to increase the capture velocity. By this the He* atoms are transversally cooled and a collimated beam of metastable helium atoms is obtained. Dimensions are not to scale, the angles are greatly exaggerated.}
\label{fig:coll}
\end{figure}

In order to intensify and collimate the atomic beam, we use two-dimensional laser cooling. The surfaces of two mutually perpendicular mirror pairs are slightly inclined with respect to the atomic beam axis (Fig.~\ref{fig:coll}). In this way, an effectively curved wavefront of the laser light is created, thereby increasing the capture angle of the collimation \cite{Aspect:1990}. Two orthogonal laser beams that are resonant with the $2^{3}\text{S}_{1} - 2^{3}\text{P}_{2}$ transition of He* enter the vacuum chamber at a capture angle of about $\SI{20}{\mrad}$ and are reflected 12 times between the mirror pairs. This geometry corresponds to a radius of curvature $R\approx\SI{7}{\m}$. The diameter of the laser beams is \SI{6}{\mm} and their intensity is $500\, I_s$ ($I_s$ is the saturation intensity of the according atomic transition).

The atomic beam passes between two metal plates kept at a potential difference of $\SI{1}{\kV}$ to deflect ions and electrons from the beam. While this was implemented as a precaution to minimize collisions of metastable atoms with charged particles in the beam, the influence on the loading rate of the trap turned out to be negligible. To optimize and measure the flux of the atomic beam, however, it is necessary to deflect charged particles from the beam as their contribution to the measurement signal can be up to an order of magnitude higher than the one from the metastable atoms. The beam can be detected either using the fluorescence of the atoms at $\SI{389}{\nm}$ ($2^{3}\text{S}_{1} - 3^{3}\text{P}_{2}$ transition) or Faraday cup detectors. The positions along the atomic beam path where we have access for fluorescence imaging are marked with \emph{FI} in Fig.~\ref{fig:overview}, the locations where our two Faraday cup detectors are installed are marked with \emph{FC}. Using collimation, we observe a 30-fold increase in the atomic flux, thereby reaching a flow of metastable atoms at \emph{FC2} exceeding $\SI{1.5e11}{\atoms/\s}$.

\section{Zeeman slower}\label{s:slower}

Even though the helium atoms are precooled in the source, the longitudinal velocity of the atomic beam exceeds by far the capture range of our magneto-optical trap. Thus, the common technique of Zeeman slowing~\cite{Phillips:1982,Dedman:2004} is used to decelerate the atoms from $v\approx\SI{800}{\m/\s}$ to less than $\SI{100}{\m/\s}$.

\begin{figure}
\centering
\includegraphics[width=\columnwidth]{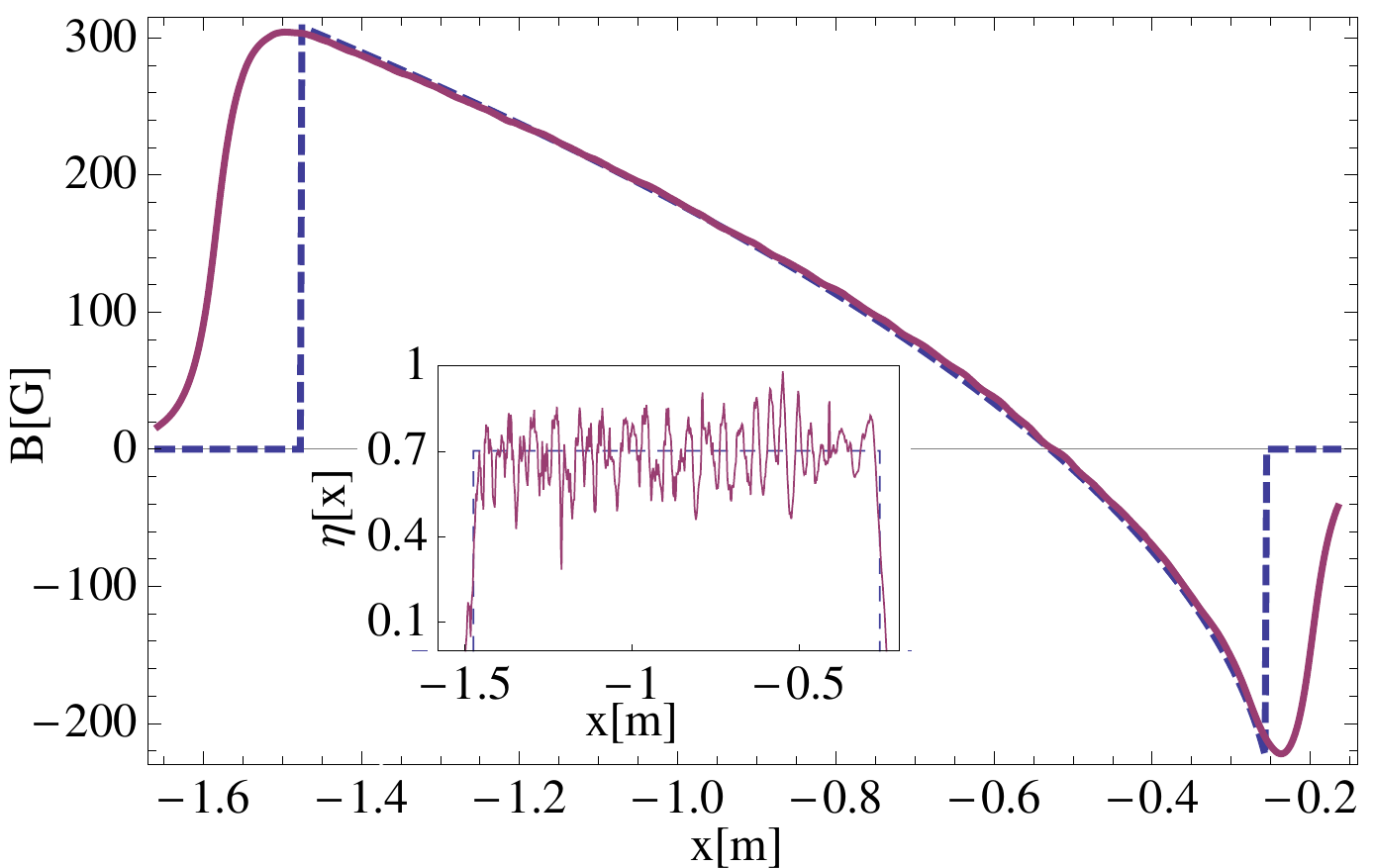}
\caption{(Color online) Target magnetic field (blue dashed) for the Zeeman slower and experimentally measured field (purple solid). The origin of the x-axis is located at the center of the atom traps in the science chamber. The inset shows the local efficiency of the slowing process. The measured efficiency (purple solid) oscillates around the theoretically expected efficiency (blue dashed) due to the discrete steps of individual layers of the solenoid.}
\label{fig:zeeman-field}
\end{figure}

The atoms are slowed by a resonant, counter-propagating laser. In order to compensate for the change in Doppler shift during deceleration, a spatially varying magnetic field is created with a tapered solenoid, which causes a position-dependent shift of the Zeeman sublevels~\cite{Phillips:1982}. To optimize the field, we model the coil system with individual windings such that the target field is matched. In total, the Zeeman slower consists of $\SI{1.7}{\km}$ enameled rectangular wire ($\SI{2.5x1.1}{\mm}$) wound onto a $\SI{1.44}{\m}$-long vacuum tube with an outer diameter of $\SI{34}{\mm}$. Seventeen double layers of windings in two sections with opposite magnetic field direction achieve a sufficiently strong magnetic field at a current of $\SI{2}{\ampere}$ without the need for active water cooling. The magnetic field profile has been measured and found to be consistent with the model. The on-axis measurement along with the target field is depicted in Fig.~\ref{fig:zeeman-field}.

While the field of an ideal Zeeman slower has a smooth variation along the axis, a real solenoid is made out of individual windings with steps at the end of each layer. At these steps, the magnetic field changes faster than what would be required for constant deceleration. To take this into account, a position-dependent efficiency $\eta(x)$ of the Zeeman slowing can be calculated~\cite{Dedman:2004},
\begin{equation}
\eta(x)=\frac{2m\mu_{\text{eff}}}{\hbar^2k_L^3\Gamma}\frac{dB(x)}{dx}\left(\frac{\mu_{\text{eff}}B(x)}{\hbar}-\delta_0\right),
\end{equation}
where $\delta_0$ is the detuning of the laser from the atomic resonance without taking the Doppler or Zeeman shift into account. As can be seen in the inset of Fig.~\ref{fig:zeeman-field}, the efficiency of the slower oscillates according to the layer structure. Once the resonance condition is lost, the atoms will not get back into resonance and are not slowed any further. It is, therefore, practical to make the slower a little longer than theoretically necessary. A length of the solenoid of $\SI{1.36}{\m}$ gives an average efficiency of $0.7$ and guarantees $\eta\leq 1$ everywhere along the trajectory of the atoms through the tube.

\begin{figure}
\begin{center}
\includegraphics[width=\columnwidth]{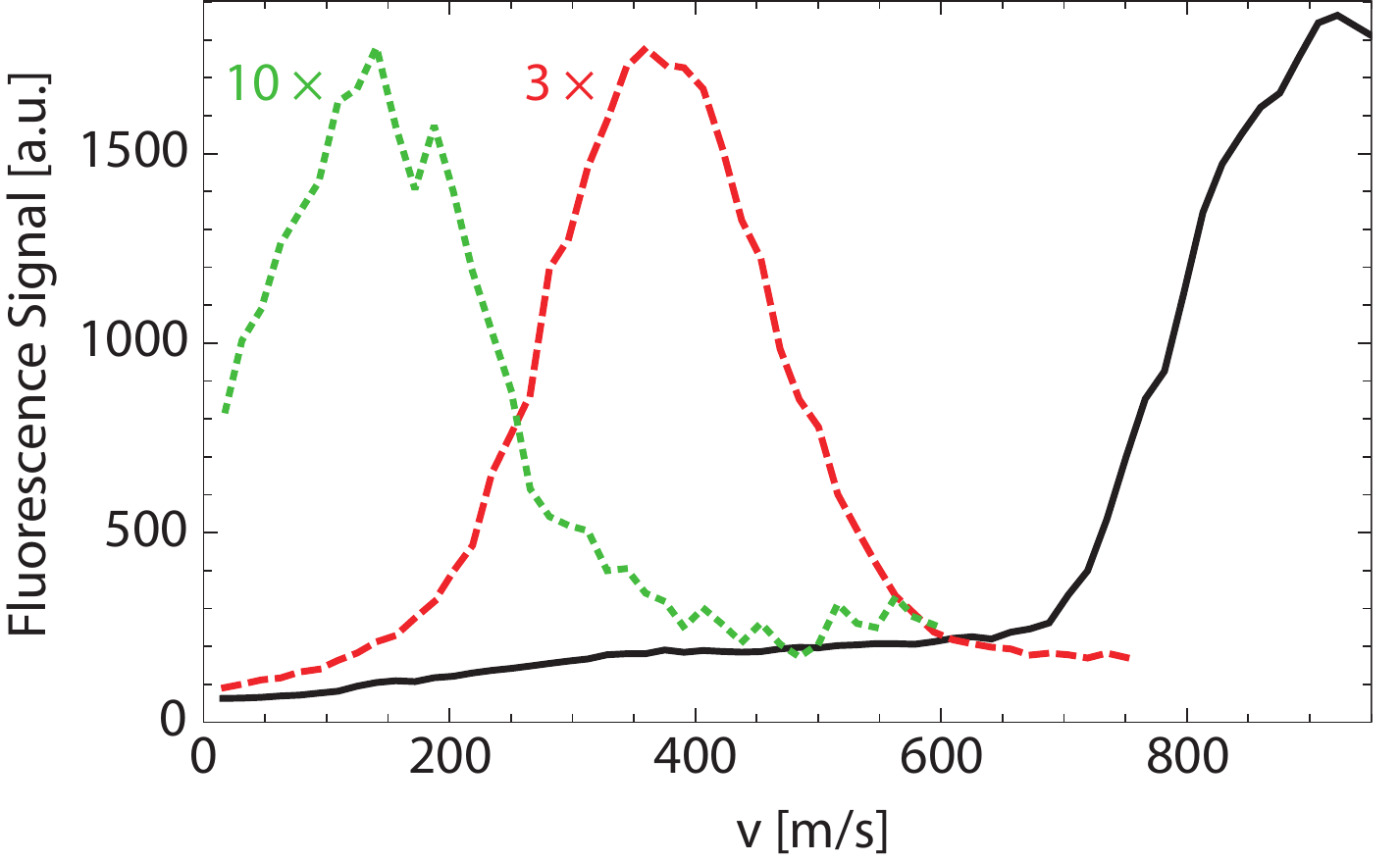}
\caption{(Color online) Example of atomic beam deceleration. The velocity of the atomic beam after the Zeeman slower is measured with a resonant laser beam that is tilted to be sensitive to the velocity component along the Zeeman-slower axis. By scanning the laser frequency we probe the velocity distribution of the atomic beam. Without Zeeman slowing, the atoms have a peak velocity of about $\SI{900}{\m/\s}$ in this measurement (black solid). The first Zeeman slower section slows the atoms down to less than $\SI{400}{\m/\s}$ (red dashed), and the second section brings the velocity down to capturable values (green dotted). To load the MOT, the field of the second stage is raised such that the atoms exiting the Zeeman slower have a velocity of about $\SI{80}{\m/\s}$. Due to transversal heating within the Zeeman slower the atomic beam expands which leads to a decrease in fluorescence signal. For better comparison of the velocity distributions the results of the slowed atoms are therefore enlarged by a factor of 3 (red dashed curve) and a factor of 10 (green dotted curve), respectively.}
\label{fig:zeeman_results}
\end{center}
\end{figure}

The solenoid for the Zeeman slower is split into two sections that create a magnetic field in two opposite directions. Figure \ref{fig:zeeman_results} shows the influence of the two slowing sections on the velocity distribution of the atomic beam. When the first solenoid is switched on, the axial velocity of the atoms is reduced in this measurement from $\SI{900}{\m/\s}$ to $\SI{400}{\m/\s}$, and the second section reduces the axial velocity further to speeds that can be captured by the MOT, typically about $\SI{80}{\m/\s}$.

In order to slow the atoms, we use similar parameters for the laser as used by other groups with He* BECs. Namely, the frequency of a $\sigma^+$ polarized laser beam at $\SI{1083}{\nm}$ is red-detuned by $\SI{-370}{\MHz}$ with respect to the unshifted atomic transition. The beam has an initial diameter of $\SI{36}{\mm}$ and is focused onto the skimmer of the source. This way the component perpendicular to the atomic beam axis counteracts the transversal heating.

\section{Magneto-optical trap}\label{s:mot}

The slowed atoms are then captured in MOT. Its center is located $\SI{20}{\cm}$ behind the Zeeman-slower exit in an ultrahigh vacuum (UHV) chamber with a base pressure in the $\SI{1e-11}{\mbar}$ regime. This pressure is achieved with a combination of turbomolecular pumps and a titanium sublimation pump.
While loading the MOT, the pressure increases to $\SI{3e-10}{\mbar}$, despite the use of a differential pumping stage ($\SI{3.5}{\mm}$ bore diameter, $\SI{150}{\mm}$ length) at the entrance of the Zeeman slower and an additional turbomolecular pump at its zero magnetic field crossing. Blocking the direct line of sight from the source to the trap center with a fast shutter and closing a vacuum gate valve after loading of the MOT allows one to perform all the following experimental steps at the base pressure of the UHV chamber.

For the MOT, we use three pairs of counterpropagating laser beams that are mutually perpendicular to each other (cf.~Fig.~\ref{fig:overview}). Their diameter is $\SI{36}{\mm}$, and their intensity is $30 \, I_s$ per beam. The light is red-detuned by $-26.5 \, \Gamma$ ($\SI{-43}{\MHz}$) from the atomic resonance. The high intensity and detuning increase the capture velocity of the MOT and keep the density of the atomic cloud low, which is necessary to minimize losses resulting from Penning ionization, a collisional process that deexcites the metastable helium atom back to its ground state and ionizes its collision partner \cite{Penning:1927}. Penning ionization can also occur between two He* atoms, in which case both atoms are lost from the trap.

The quadrupole magnetic field for the MOT is generated by the outer radial (OR) coils that are part of the ``cloverleaf'' magnetic trap assembly (Fig.~\ref{fig:mt}). The coils have a diameter of $\SI{12}{\cm}$ and are $\SI{10}{\cm}$ apart. The magnetic field gradient at the trap center is $\SI{20}{\gauss/\cm}$ in radial and $\SI{10}{\gauss/\cm}$ in axial direction with respect to the symmetry axis of the magnetic trap.

We trap $\num{6e8}$ atoms at a temperature of $\SI{1.5}{\mK}$. The lifetime of an atom cloud in our MOT is $\SI{190}{\ms}$ and it has a radius of $\SI{1.9}{\mm}$. Those numbers are comparable to other setups of metastable helium BECs.

\section{Magnetic trap}\label{s:mt}

\begin{figure}
\begin{center}
\includegraphics[width=\columnwidth]{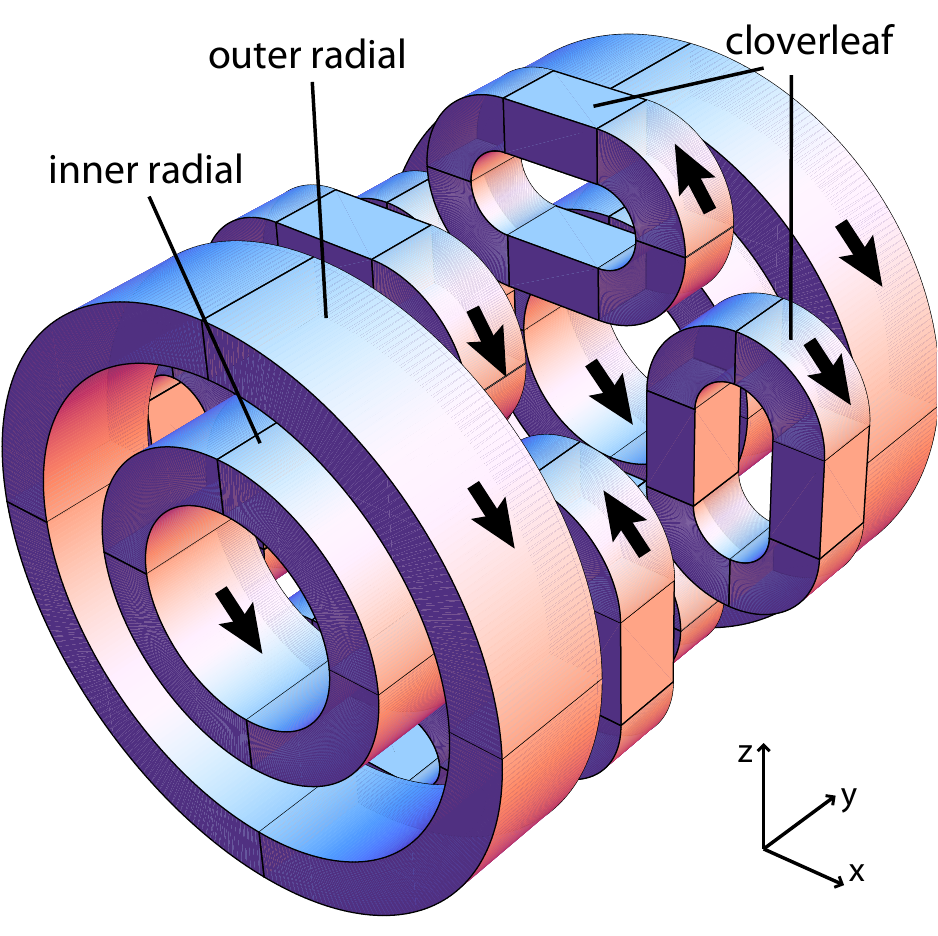}
\caption{\label{fig:mt}(Color online) Schematic arrangement of our ``cloverleaf'' magnetic trap. It consists of a pair of outer radial coils (OR), a pair of inner radial coils (IR), and four pairs of racetrack-shaped ``cloverleaf'' coils (CL). The MOT quadrupole field is created by the current flowing in opposite directions in the OR coils, whereas in the magnetic trap OR and IR coils are mainly responsible for the trap bias field and the axial confinement, and the CL coils create the radial confinement. The coils are wound out of square, water-cooled, copper tubing. Each coil consists of three layers, each containing six windings of copper tubing, except for CL coils that have three windings in each layer. The arrows indicate current directions in the magnetic trap.}
\end{center}
\end{figure}

The quadrupole potential of the magnetic field in the MOT, due to its zero crossing, is not suited for trapping atoms for longer times. Therefore, one has to create a potential that has a nonzero magnetic field minimum. We have constructed a ``cloverleaf'' magnetic trap \cite{Pritchard:1983}, depicted in Fig.~\ref{fig:mt}, which gives $\SI{360}{\degree}$ optical access around the symmetry axis of the coil arrangement. The minimum of the magnetic field is located at the MOT center.

To load the atoms into the magnetic trap, we need to reverse the current in one of the OR coils with respect to the MOT configuration. We switch off the MOT current leading to a decaying magnetic field over $\SI{2}{\ms}$. Meanwhile, we apply a $\SI{0.3}{\ms}$ three-dimensional optical molasses pulse with a detuning of $-2.5\,\Gamma$ ($\SI{-4}{\MHz}$) and an intensity of $5\, I_s$. This reduces the temperature of the cloud further to $\SI{1}{\mK}$ and increases the density of the atomic cloud by more than an order of magnitude. Further cooling is avoided at this point to keep the density of the atomic cloud low enough and thereby losses due to Penning ionization to a minimum.

Immediately after the molasses pulse, we switch on the currents of the cloverleaf magnetic trap, with its potential mode matched to the MOT. Simultaneously, two counterpropagating laser beams, aligned along the quantization axis of the trap, spin polarize the atomic sample  within $\SI{0.5}{\ms}$ to the magnetic low-field seeking state. Even though the beams are blue-detuned by $\SI{32}{\MHz}$ ($20\,\Gamma$) with respect to the unshifted atomic transition, they become resonant during the build-up of the magnetic field. Thus, the exact detuning as well as the laser intensity ($2\,I_{s}$ in our case) are not crucial and can be adjusted over a broad range. As has been shown~\cite{Herschbach:2000}, the spin polarization not only increases the number of atoms in the trap but also significantly reduces two-body losses due to Penning ionization. We have measured a threefold increase in the number of atoms in the trap due to the optical pumping of atoms to the low-field seeking state.

At this point, the magnetic trap has trapping frequencies of $\omega_{\text{rad}}/2\pi=\SI{32}{\Hz}$ and $\omega_{\text{ax}}/2\pi=\SI{62}{\Hz}$ in radial and axial direction, respectively, and the bias field is $\SI{24}{\G}$. We transfer about $90\%$ of the initial atoms into the magnetic trap where they are in the $2^{3}$S$_{1}$, $m_{J}=+1$ magnetic substate. The trap depth is around $\SI{2}{\mK}$. With the gate valve in front of the Zeeman slower closed, we achieve a lifetime of $\SI{61}{\s}$, which is limited by collisions with background gas.

Subsequently, we apply one-dimensional laser cooling in the magnetic trap \cite{Hijmans:1989,Schmidt:2003} for $\SI{1.2}{\s}$, which reduces the temperature of the sample to $\SI{150}{\uK}$. Two counterpropagating beams are detuned by  $-5\,\Gamma$ ($-\SI{8}{\MHz}$)  with respect to the  Zeeman-shifted transition at the trap minimum. They are $\sigma^{+}$ polarized and have an intensity of $\num{1e-4}\, I_s$. In order to increase the elastic collision rate, we increase the density of the atoms by adiabatic compression of the trap to $\omega_{\text{rad}}/2\pi=\SI{800(13)}{\Hz}$ and $\omega_{\text{ax}}/2\pi=\SI{47\pm5}{Hz}$ in the radial and axial direction, respectively, which corresponds to a radial gradient of $\SI{95}{\G\per\cm}$, an axial curvature of $\SI{11}{\G\per\cm\squared}$ and a bias field of $\SI{2}{\G}$. At this point, we have $\num{5e8}$ atoms at a phase space density of $n{\lambda_{\text{dB}}^{3}}=\num{1e-6}$ (where the density of the atomic cloud $n$ and the thermal de Broglie wavelength $\lambda_{\text{dB}}$ can be obtained from absorption measurements), which are excellent conditions to start forced evaporative cooling.

Stability of the currents in the magnetic trap coils is crucial for limiting the heating rate and thereby improving the lifetime of the atomic cloud. Also the repeatability of the magnetic field distribution when performing the same experiment multiple times is essential, especially for experiments that require good statistics. Particularly the magnitude of the magnetic field at the trap center is very susceptible to fluctuations since it results from the subtraction of the much larger magnetic fields created by the outer radial coils and the inner radial coils. To achieve those two stability requirements, the currents are measured with the help of precise current transducers and shunt resistors with a very low susceptibility to temperature changes and aging effects. The measured currents are then stabilized by analog proportional-integral-derivative (PID) control circuits to adjustable reference voltages from low-noise battery sources. The remaining rms noise can be deduced by measuring the fluctuations on the error signal of the PID circuits, which is around $\SI{1e-4}{}$. This translates to a $1\%$ stability of the $\SI{2}{\G}$ magnetic field at the trap center since it results from the subtraction of two fields on the order of $\SI{160}{\G}$.

\section{Bose-Einstein condensation}\label{s:bec}

\begin{figure}
\begin{center}
\includegraphics[width=\columnwidth]{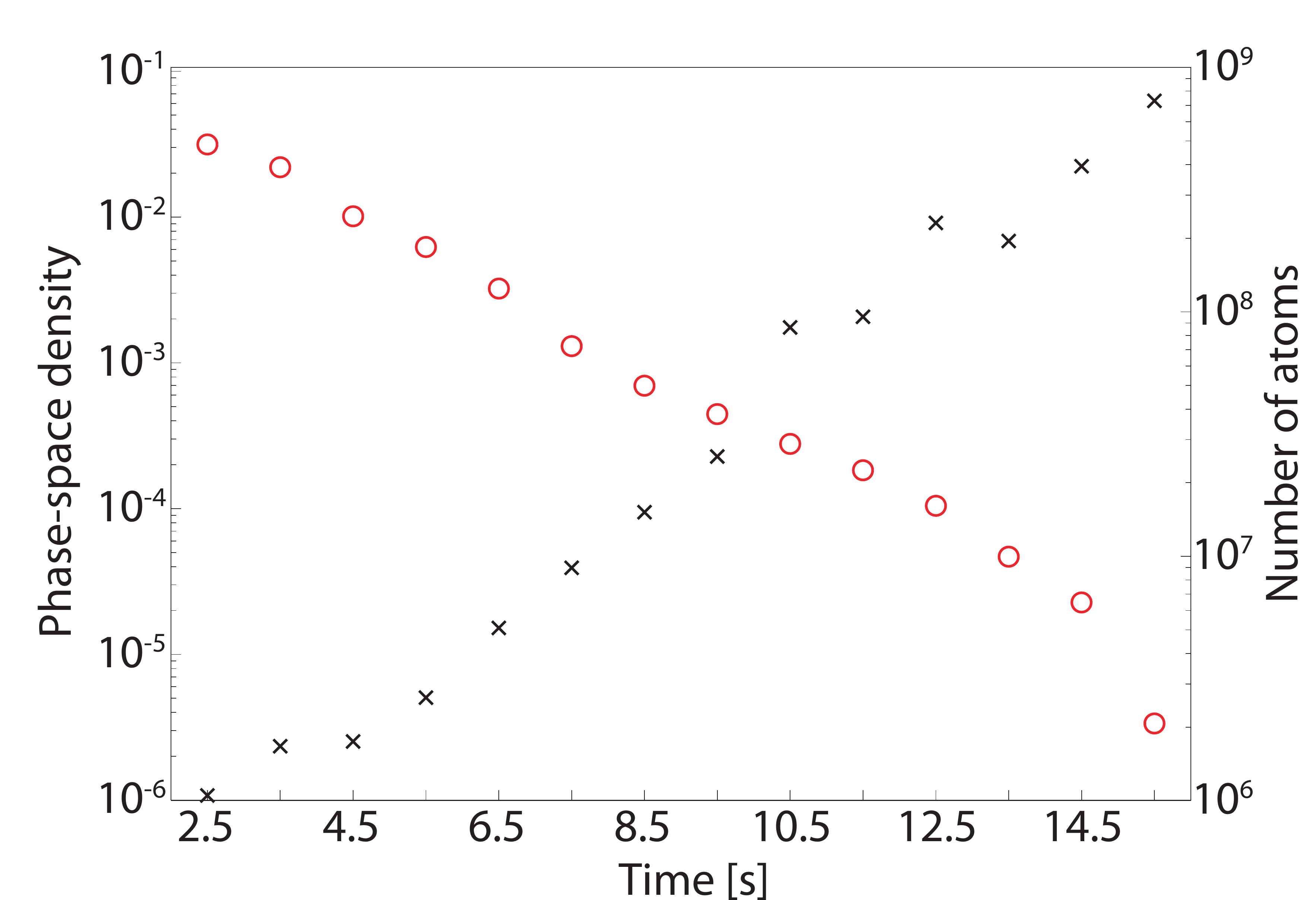}

\caption{\label{fig:rf}(Color online) Increase of the phase-space density (black crosses) and decrease of the number of atoms (red circles) during the evaporative cooling. We lose more than $\SI{99}{\%}$ of the atoms, but the phase-space density increases by five orders of magnitude. While in this realization we evaporated the atoms over a time span of $\SI{16}{\s}$, degeneracy of the atoms can be achieved in less than half of this time, which only slightly decreases the remaining number of atoms.}
\end{center}
\end{figure}

Further cooling of the atoms is done by radio-frequency-induced (rf) evaporative cooling, where the atoms with the highest energy are spin-flipped and become untrapped, whereas the remaining atoms rethermalize at a lower temperature \cite{Ketterle:1996,Luiten:1996,Herschbach:2003}. We evaporate atoms by applying up to $\SI{20}{\W}$ of rf power to a pair of coil antennas and ramping the frequency exponentially from $\SI{75}{\MHz}$ to below $\SI{6}{\MHz}$ over a time period of $\SI{8}{\s}$. At the end of the ramp, typically a few $\num{1e6}$ atoms undergo a phase transition to the BEC. Shorter ramps also lead to a BEC, albeit with smaller numbers of atoms. Figure~\ref{fig:rf} shows the increase of phase-space density at the cost of losing atoms over the duration of the rf ramp.

\begin{figure}
\begin{center}
\includegraphics[width=\columnwidth]{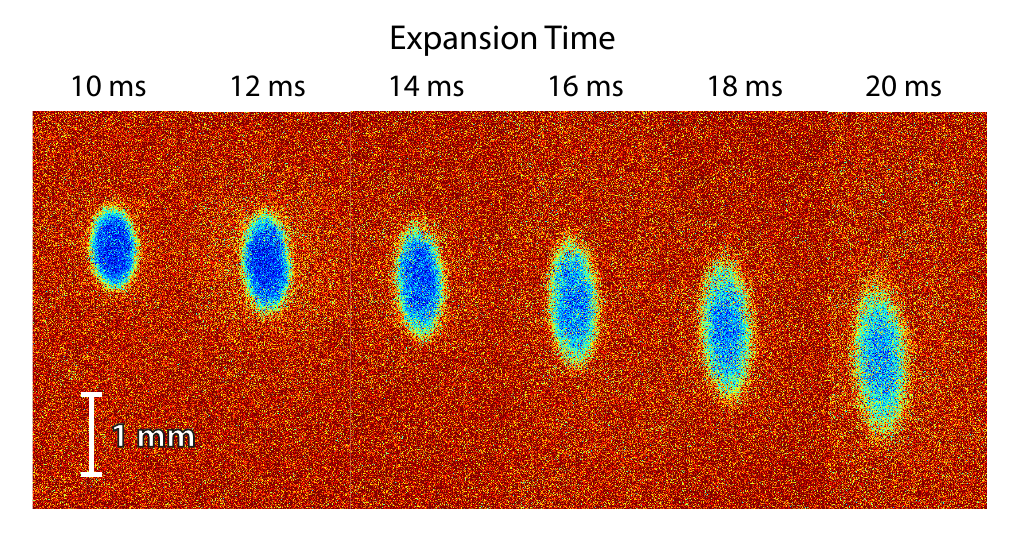}
\caption{\label{fig:bec_tof}(Color online) Absorption images of the condensed cloud falling under gravity. The observed anisotropic expansion is a signature of a BEC.}
\end{center}
\end{figure}

\begin{figure}
\begin{center}
\includegraphics[width=\columnwidth]{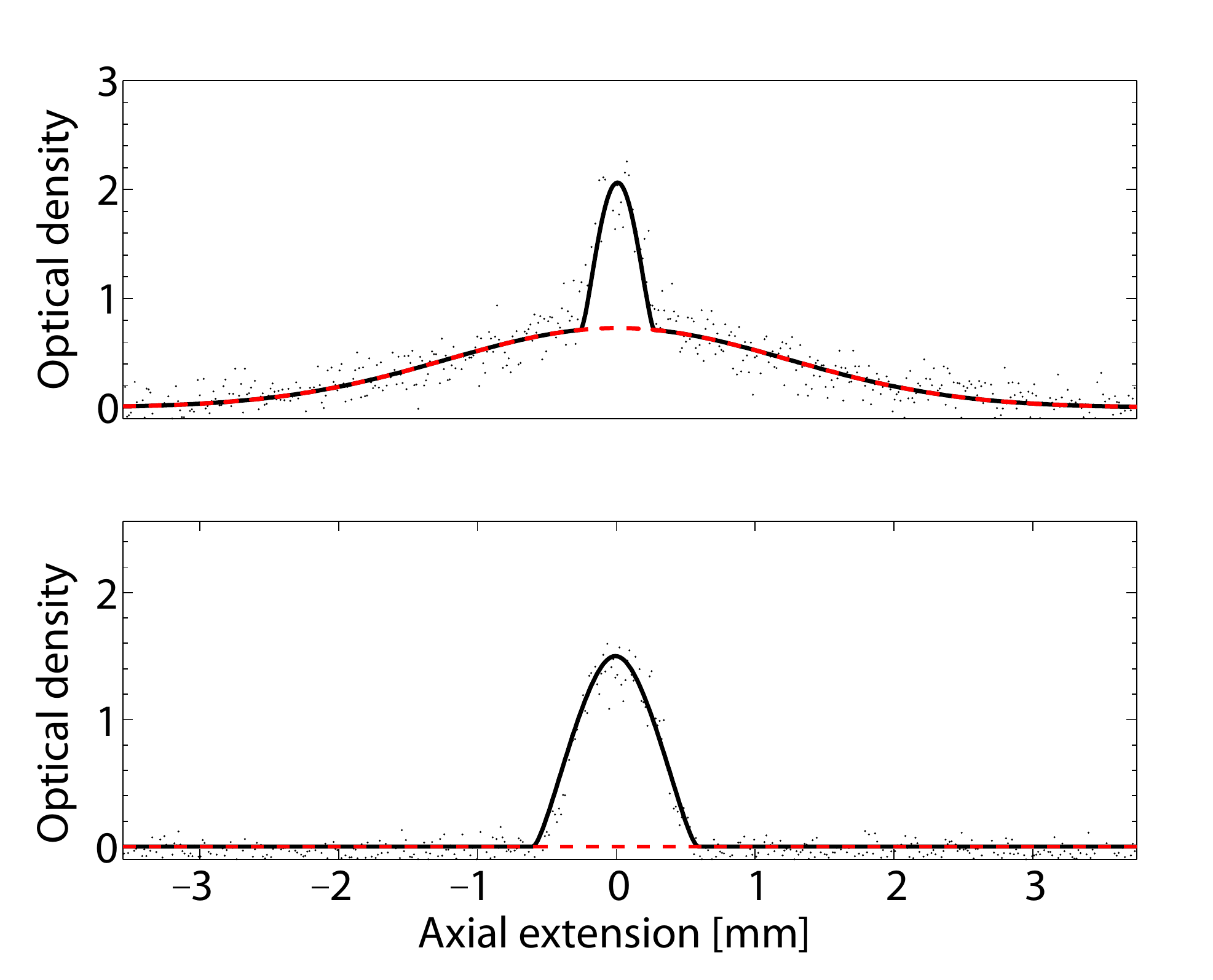}
\caption{\label{fig:bec_cross}(Color online) Density profiles (10 pixel line average) of a partly condensed atomic cloud (top picture) and a BEC (bottom picture), obtained by using different end frequencies of the rf ramp. The images were taken $\SI{15}{\ms}$ after release of the atoms from the trap. The black solid line represents the fit of the bimodal distribution, and the red dashed line the thermal component, from which the temperature of the cloud can be inferred.}
\end{center}
\end{figure}

A signature of boson degeneracy is  the asymmetric expansion after release from the trap (Fig.~\ref{fig:bec_tof}), as well as a parabolic density distribution (Fig.~\ref{fig:bec_cross}). From the Thomas-Fermi approximation one can estimate the initial size of the condensate; we find radial and axial radii of $R_{\text{rad}}\approx\SI{7}{\um}$ and $R_{\text{ax}}\approx\SI{140}{\um}$, respectively. The lifetime of the BEC is $\SI{250}{\ms}$.

\section{Detection of metastable atoms}\label{s:detection}

We have two means of detecting and analyzing the cold atomic cloud: by absorption imaging and by using a delay-line detector (DLD).

\begin{figure}
\begin{center}
\includegraphics[width=\columnwidth]{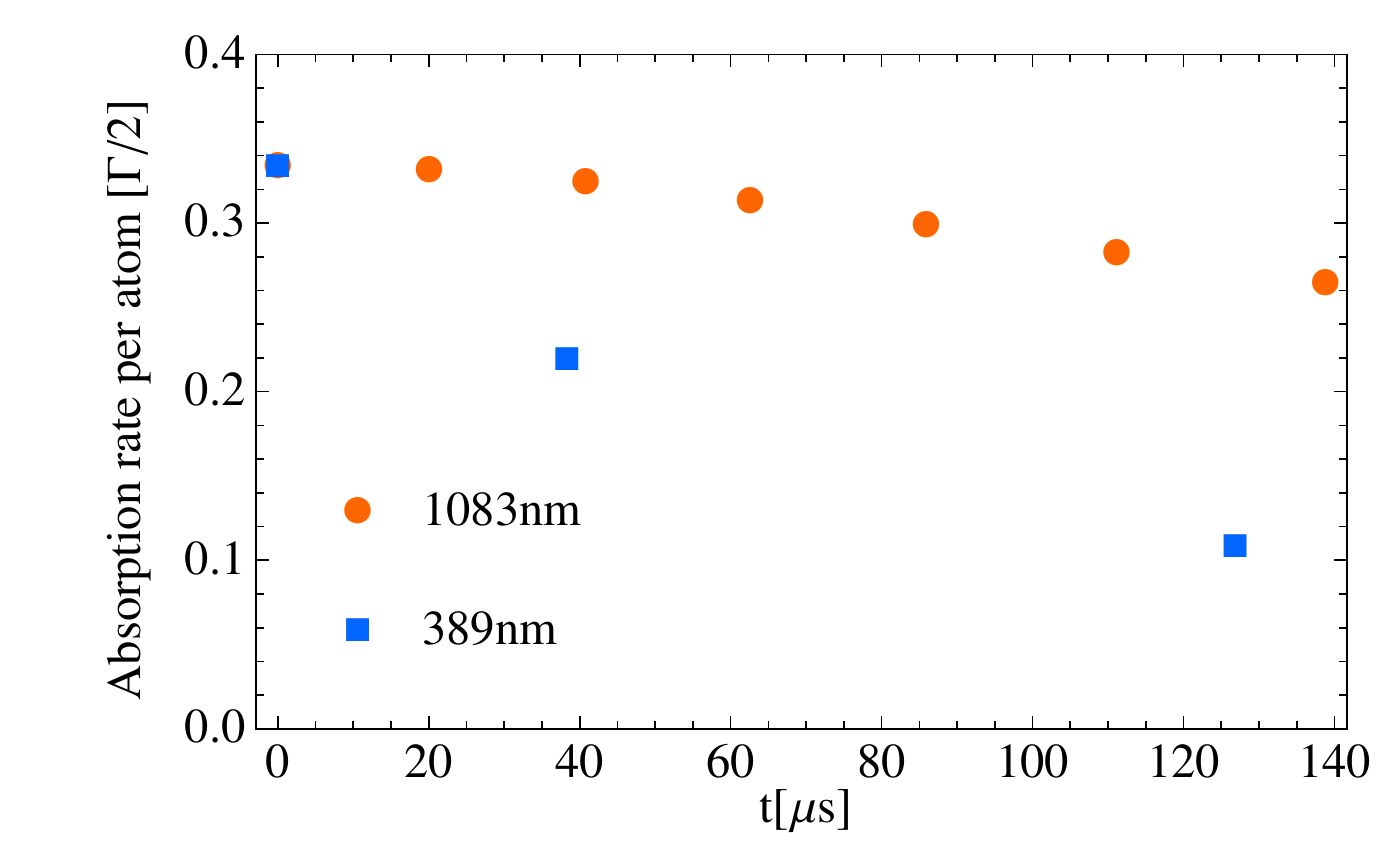}
\caption{(Color online) Absorption rate per atom over time in units of the maximum possible rate $\Gamma / 2$ at an intensity of $0.5\,I_s$. Due to the higher photon recoil for $\SI{389}{\nm}$ photons (blue squares), the absorption rate drops much faster than for $\SI{1083}{\nm}$ photons (red circles). The horizontal spacing between the dots indicates the average time between the scattering of two photons~\cite{Herschbach:2003:thesis}. Therefore, by using the $\SI{1083}{\nm}$ transition for absorption imaging of the atomic cloud, the atoms stay resonant and scatter photons for a longer time than in the case of $\SI{389}{\nm}$ laser light.}
\label{fig:scatteringrate}
\end{center}
\end{figure}

Absorption imaging is used to detect and analyze the whole ensemble of atoms in the cold atomic cloud. An EMCCD camera delivers high detection efficiency at $\SI{389}{\nm}$, which can be exploited for fluorescence imaging of the atomic beam. However, for the cold cloud with a limited number of atoms, the large momentum transfer of the blue light rapidly accelerates the atoms out of resonance with the imaging beam (compare Fig.~\ref{fig:scatteringrate}). Technical limitations with the drivers of our acousto-optical modulators set a lower limit on the length of the imaging laser pulses to about $\SI{50}{\us}$. If the atoms are pushed out of resonance on a shorter time scale, the contrast in the image decreases which cannot be compensated for by a high detection efficiency. Instead we use $0.01\,I_{s}$ of resonant light at $\SI{1083}{\nm}$ and image the shadow of the cloud. From the optical depth of the atomic cloud we deduce the number of atoms and from time-of-flight measurements obtain the temperature of the cloud.

\begin{figure}
\begin{center}
\includegraphics[width=\columnwidth]{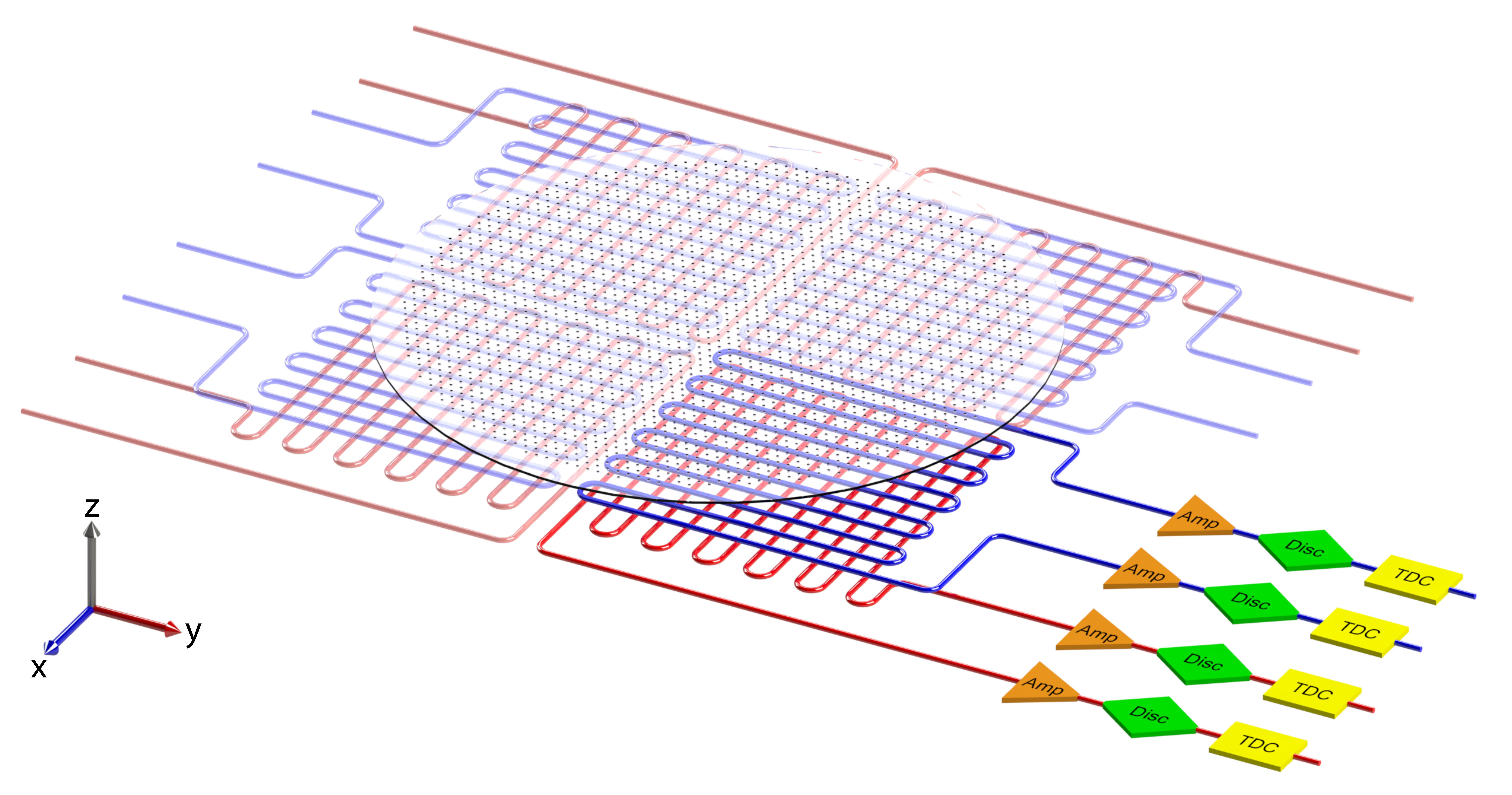}
\caption{\label{fig:dld}(Color online) Schematic of the delay-line detector (DLD) consisting of four quadrants with independent delay lines and readout electronics. Metastable helium atoms falling onto the micro-channel plate (MCP) get deexcited and are likely to cause the emission of an electron that can be amplified to an electron avalanche in one of the channels. When it hits the delay lines located underneath the MCP, it creates an electronic signal that propagates along the wires in both directions. After preamplification, pulse discrimination, and time-to-digital conversion (TDC), the four arrival times are recorded. Due to a constant propagation speed of the electronic signals along the delay lines, the arrival position of the helium atom can be deduced from $t_{x1}-t_{x2}$ and $t_{y1}-t_{y2}$. The arrival time of the atom at the detector is determined by taking the average of the four recorded times. The active region of the MCP has a diameter of $\SI{80}{\mm}$; the delay lines of each quadrant extend over rectangular regions with a size of $\SI{45x48}{\mm}$ with a separation between two neighboring quadrants of below $\SI{1}{\mm}$. At the turning points of the delay lines strong distortions are introduced that cannot be corrected for by calibrating the detector. Therefore a region of approximately $\SI{5}{\mm}$ in between the quadrants has to be discarded.}
\end{center}
\end{figure}

The key element in our setup, though, is the DLD (Fig.~\ref{fig:dld}), which allows for single-atom detection. It is mounted $\SI{80}{\cm}$ below the trap and can be moved and rotated in the $xy$-plane by translational and rotational stages, respectively. The detection process starts at a micro-channel plate (MCP), which emits an avalanche of electrons when hit by a highly energetic particle. Subsequent recording of the arrival times of the electronic signal propagating along a delay line reveals its impact position and time.

Due to the propagation of the electronic signal along the delay lines, a dead time of $\SI{25}{\ns}$ between two recordable events arises. Besides the dead time, electronic read-out and charge depletion of the MCP itself limit the burst count rate to a few $\SI{1e6}{\per\second}$. The specialty of our DLD design is having the delay line itself split into four quadrants, each equipped with independent read-out electronics. This overcomes the dead time for particles not falling onto the same quadrant. Hence, true simultaneous detection of atoms registered at different quadrants can be achieved, which will be especially useful for quantum correlation experiments with a directed emission of particles.

\begin{figure}
\begin{center}
\includegraphics[width=\columnwidth]{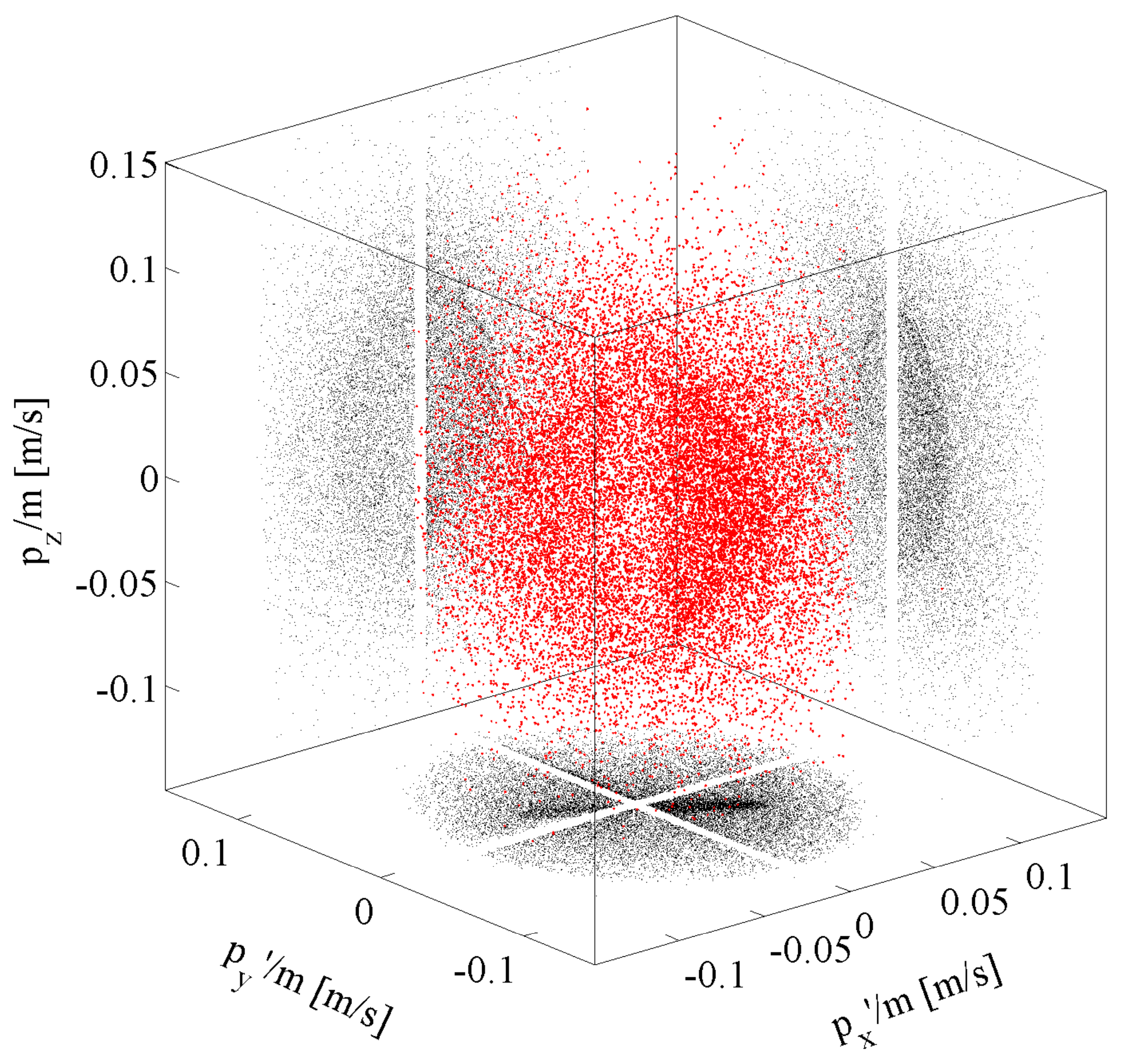}
\caption{(Color online) Reconstructed momentum space distribution of an atomic cloud that was dropped onto the DLD, where each dot represents a single atom (in total about $\num{3e4}$ for this measurement). Knowing the distance of the trap to the detector, the momentum of the atoms can be reconstructed. While no specific structure can be seen by the bare eye in the three-dimensional plot, the two-dimensional projections show the higher density of the condensed part of the atoms. The four different quadrants of the detector can be identified in the projections because of the areas in between the quadrants where data has been discarded due to non-correctable distortions. Note that the detector was rotated with respect to the magnetic-trap symmetry axis in this measurement, therefore $x\prime \neq x$, $y\prime \neq y$.}
\label{fig:reconstruct_bec}
\end{center}
\end{figure}

We measure a temporal precision of $\SI{220\pm18}{\ps}$ and a spatial precision of $\SI{177\pm32}{\um}$. These values represent rms deviations from the mean values. The measurement of the temporal resolution has been performed by shining a picosecond mode-locked UV laser on the quadrants of the detector and analyzing the fluctuations of the registered detection events around the laser repetition rate. The spatial resolution has been directly measured by placing a mask with $\SI{30}{\um}$-wide holes on top of the MCP and deconvoluting the obtained pattern with the size of the holes. The obtained values are much larger than the time-bin size of the time-to-digital converter ($\SI{6.8}{\ps}$) or the spatial discretization of our detector (about $\SI{30}{\um}$) and represent the actual performance in our current setup. Including the filling factor of the MCP, we estimate the efficiency of the overall process for single metastable helium atoms to be approximately $\SI{7}{\%}$ by comparing with numbers of atoms measured in absorption imaging. The estimate is on the conservative side and compares well to what has been reported by the Palaiseau group~\cite{Perrin:2007}. From the spatio-temporal information of the detection event, the momentum-space distribution of the atoms can be reconstructed (Fig.~\ref{fig:reconstruct_bec}), which will be an essential element for future correlation experiments.

\section{Conclusions}

To achieve Bose-Einstein condensation, a sequence of well-controlled manipulation and cooling steps has been designed, developed, implemented and characterized. First, the atoms are excited to their metastable state within a dc-discharge source. Transverse laser cooling increases the atomic beam flux, which is necessary to achieve high loading rates into a magneto-optical trap. The MOT is loaded after deceleration of the atoms in a Zeeman slower. Following the transfer to a magnetic trap, the atoms are cooled to degeneracy, with typically a few $\num{1e6}$ atoms at temperatures of around $\SI{1}{\uK}$. This will eventually serve as a source for coherent matter-wave experiments.

Moreover, a micro-channel plate in combination with a delay-line detector has been attached to the system and its characteristics were tested. The delay-line detector has the capability to resolve single-atom detection events temporally with $\SI{220\pm18}{\ps}$ and spatially with $\SI{177\pm32}{\um}$ at rates of several $\num{1e6}$ events per second and quadrant. 

The combination of a BEC as source for coherent matter waves and single-atom detection capability to measure correlations will enable us to perform experiments with non-classical states of matter waves that can be created via four-wave mixing in the condensate. The combination of high count rates, good resolution, and true simultaneous detection with independent quadrants of the delay-line detector provides excellent conditions for this type of experiments. It will bring a three-dimensional version of the original Einstein-Podolsky-Rosen paradox closer to realization in a matter-wave system and push the field closer to exploiting the possibilities of matter-wave quantum entanglement.

\begin{acknowledgments}
We like to thank Jook Walraven for comments on our manuscript.

The research was funded by the Doctoral Program CoQuS (Grant No. {W~1210}) and {SFB~F~4007} of the Austrian Science Fund (FWF).
\end{acknowledgments}

\bibliographystyle{apsrev4-1}
\bibliography{references}

\begin{thebibliography}{48}%
\makeatletter
\providecommand \@ifxundefined [1]{%
 \@ifx{#1\undefined}
}%
\providecommand \@ifnum [1]{%
 \ifnum #1\expandafter \@firstoftwo
 \else \expandafter \@secondoftwo
 \fi
}%
\providecommand \@ifx [1]{%
 \ifx #1\expandafter \@firstoftwo
 \else \expandafter \@secondoftwo
 \fi
}%
\providecommand \natexlab [1]{#1}%
\providecommand \enquote  [1]{``#1''}%
\providecommand \bibnamefont  [1]{#1}%
\providecommand \bibfnamefont [1]{#1}%
\providecommand \citenamefont [1]{#1}%
\providecommand \href@noop [0]{\@secondoftwo}%
\providecommand \href [0]{\begingroup \@sanitize@url \@href}%
\providecommand \@href[1]{\@@startlink{#1}\@@href}%
\providecommand \@@href[1]{\endgroup#1\@@endlink}%
\providecommand \@sanitize@url [0]{\catcode `\\12\catcode `\$12\catcode
  `\&12\catcode `\#12\catcode `\^12\catcode `\_12\catcode `\%12\relax}%
\providecommand \@@startlink[1]{}%
\providecommand \@@endlink[0]{}%
\providecommand \url  [0]{\begingroup\@sanitize@url \@url }%
\providecommand \@url [1]{\endgroup\@href {#1}{\urlprefix }}%
\providecommand \urlprefix  [0]{URL }%
\providecommand \Eprint [0]{\href }%
\providecommand \doibase [0]{http://dx.doi.org/}%
\providecommand \selectlanguage [0]{\@gobble}%
\providecommand \bibinfo  [0]{\@secondoftwo}%
\providecommand \bibfield  [0]{\@secondoftwo}%
\providecommand \translation [1]{[#1]}%
\providecommand \BibitemOpen [0]{}%
\providecommand \bibitemStop [0]{}%
\providecommand \bibitemNoStop [0]{.\EOS\space}%
\providecommand \EOS [0]{\spacefactor3000\relax}%
\providecommand \BibitemShut  [1]{\csname bibitem#1\endcsname}%
\let\auto@bib@innerbib\@empty
\bibitem [{\citenamefont {Anderson}\ \emph {et~al.}(1995)\citenamefont
  {Anderson}, \citenamefont {Ensher}, \citenamefont {Matthews}, \citenamefont
  {Wieman},\ and\ \citenamefont {Cornell}}]{Anderson:1995}%
  \BibitemOpen
  \bibfield  {author} {\bibinfo {author} {\bibfnamefont {M.~H.}\ \bibnamefont
  {Anderson}}, \bibinfo {author} {\bibfnamefont {J.~R.}\ \bibnamefont
  {Ensher}}, \bibinfo {author} {\bibfnamefont {M.~R.}\ \bibnamefont
  {Matthews}}, \bibinfo {author} {\bibfnamefont {C.~E.}\ \bibnamefont
  {Wieman}}, \ and\ \bibinfo {author} {\bibfnamefont {E.~A.}\ \bibnamefont
  {Cornell}},\ }\href@noop {} {\bibfield  {journal} {\bibinfo  {journal}
  {Science}\ }\textbf {\bibinfo {volume} {269}},\ \bibinfo {pages} {198}
  (\bibinfo {year} {1995})}\BibitemShut {NoStop}%
\bibitem [{\citenamefont {Davis}\ \emph {et~al.}(1995)\citenamefont {Davis},
  \citenamefont {Mewes}, \citenamefont {Andrews}, \citenamefont {van Druten},
  \citenamefont {Durfee}, \citenamefont {Kurn},\ and\ \citenamefont
  {Ketterle}}]{Davis:1995}%
  \BibitemOpen
  \bibfield  {author} {\bibinfo {author} {\bibfnamefont {K.~B.}\ \bibnamefont
  {Davis}}, \bibinfo {author} {\bibfnamefont {M.-O.}\ \bibnamefont {Mewes}},
  \bibinfo {author} {\bibfnamefont {M.~R.}\ \bibnamefont {Andrews}}, \bibinfo
  {author} {\bibfnamefont {N.~J.}\ \bibnamefont {van Druten}}, \bibinfo
  {author} {\bibfnamefont {D.~S.}\ \bibnamefont {Durfee}}, \bibinfo {author}
  {\bibfnamefont {D.~M.}\ \bibnamefont {Kurn}}, \ and\ \bibinfo {author}
  {\bibfnamefont {W.}~\bibnamefont {Ketterle}},\ }\href@noop {} {\bibfield
  {journal} {\bibinfo  {journal} {Physical Review Letters}\ }\textbf {\bibinfo
  {volume} {75}},\ \bibinfo {pages} {3969} (\bibinfo {year}
  {1995})}\BibitemShut {NoStop}%
\bibitem [{\citenamefont {Bradley}\ \emph {et~al.}(1995)\citenamefont
  {Bradley}, \citenamefont {Sackett}, \citenamefont {Tollett},\ and\
  \citenamefont {Hulet}}]{Bradley:1995}%
  \BibitemOpen
  \bibfield  {author} {\bibinfo {author} {\bibfnamefont {C.~C.}\ \bibnamefont
  {Bradley}}, \bibinfo {author} {\bibfnamefont {C.~A.}\ \bibnamefont
  {Sackett}}, \bibinfo {author} {\bibfnamefont {J.~J.}\ \bibnamefont
  {Tollett}}, \ and\ \bibinfo {author} {\bibfnamefont {R.~G.}\ \bibnamefont
  {Hulet}},\ }\href@noop {} {\bibfield  {journal} {\bibinfo  {journal}
  {Physical Review Letters}\ }\textbf {\bibinfo {volume} {75}},\ \bibinfo
  {pages} {1687} (\bibinfo {year} {1995})}\BibitemShut {NoStop}%
\bibitem [{\citenamefont {Greiner}\ \emph {et~al.}(2002)\citenamefont
  {Greiner}, \citenamefont {Mandel}, \citenamefont {Esslinger}, \citenamefont
  {H{\"a}nsch},\ and\ \citenamefont {Bloch}}]{Greiner:2002}%
  \BibitemOpen
  \bibfield  {author} {\bibinfo {author} {\bibfnamefont {M.}~\bibnamefont
  {Greiner}}, \bibinfo {author} {\bibfnamefont {O.}~\bibnamefont {Mandel}},
  \bibinfo {author} {\bibfnamefont {T.}~\bibnamefont {Esslinger}}, \bibinfo
  {author} {\bibfnamefont {T.~W.}\ \bibnamefont {H{\"a}nsch}}, \ and\ \bibinfo
  {author} {\bibfnamefont {I.}~\bibnamefont {Bloch}},\ }\href@noop {}
  {\bibfield  {journal} {\bibinfo  {journal} {Nature}\ }\textbf {\bibinfo
  {volume} {415}},\ \bibinfo {pages} {39} (\bibinfo {year} {2002})}\BibitemShut
  {NoStop}%
\bibitem [{\citenamefont {Jochim}\ \emph {et~al.}(2003)\citenamefont {Jochim},
  \citenamefont {Bartenstein}, \citenamefont {Altmeyer}, \citenamefont {Hendl},
  \citenamefont {Riedl}, \citenamefont {Chin}, \citenamefont
  {Hecker~Denschlag},\ and\ \citenamefont {Grimm}}]{Jochim:2003}%
  \BibitemOpen
  \bibfield  {author} {\bibinfo {author} {\bibfnamefont {S.}~\bibnamefont
  {Jochim}}, \bibinfo {author} {\bibfnamefont {M.}~\bibnamefont {Bartenstein}},
  \bibinfo {author} {\bibfnamefont {A.}~\bibnamefont {Altmeyer}}, \bibinfo
  {author} {\bibfnamefont {G.}~\bibnamefont {Hendl}}, \bibinfo {author}
  {\bibfnamefont {S.}~\bibnamefont {Riedl}}, \bibinfo {author} {\bibfnamefont
  {C.}~\bibnamefont {Chin}}, \bibinfo {author} {\bibfnamefont {J.}~\bibnamefont
  {Hecker~Denschlag}}, \ and\ \bibinfo {author} {\bibfnamefont
  {R.}~\bibnamefont {Grimm}},\ }\href@noop {} {\bibfield  {journal} {\bibinfo
  {journal} {Science}\ }\textbf {\bibinfo {volume} {302}},\ \bibinfo {pages}
  {2101} (\bibinfo {year} {2003})}\BibitemShut {NoStop}%
\bibitem [{\citenamefont {Zwierlein}\ \emph {et~al.}(2003)\citenamefont
  {Zwierlein}, \citenamefont {Stan}, \citenamefont {Schunck}, \citenamefont
  {Raupach}, \citenamefont {Gupta}, \citenamefont {Hadzibabic},\ and\
  \citenamefont {Ketterle}}]{Zwierlein:2003}%
  \BibitemOpen
  \bibfield  {author} {\bibinfo {author} {\bibfnamefont {M.~W.}\ \bibnamefont
  {Zwierlein}}, \bibinfo {author} {\bibfnamefont {C.~A.}\ \bibnamefont {Stan}},
  \bibinfo {author} {\bibfnamefont {C.~H.}\ \bibnamefont {Schunck}}, \bibinfo
  {author} {\bibfnamefont {S.~M.~F.}\ \bibnamefont {Raupach}}, \bibinfo
  {author} {\bibfnamefont {S.}~\bibnamefont {Gupta}}, \bibinfo {author}
  {\bibfnamefont {Z.}~\bibnamefont {Hadzibabic}}, \ and\ \bibinfo {author}
  {\bibfnamefont {W.}~\bibnamefont {Ketterle}},\ }\href@noop {} {\bibfield
  {journal} {\bibinfo  {journal} {Physical Review Letters}\ }\textbf {\bibinfo
  {volume} {91}},\ \bibinfo {pages} {250401} (\bibinfo {year}
  {2003})}\BibitemShut {NoStop}%
\bibitem [{\citenamefont {Greiner}\ \emph {et~al.}(2003)\citenamefont
  {Greiner}, \citenamefont {Regal},\ and\ \citenamefont {Jin}}]{Greiner:2003}%
  \BibitemOpen
  \bibfield  {author} {\bibinfo {author} {\bibfnamefont {M.}~\bibnamefont
  {Greiner}}, \bibinfo {author} {\bibfnamefont {C.~A.}\ \bibnamefont {Regal}},
  \ and\ \bibinfo {author} {\bibfnamefont {D.~S.}\ \bibnamefont {Jin}},\
  }\href@noop {} {\bibfield  {journal} {\bibinfo  {journal} {Nature}\ }\textbf
  {\bibinfo {volume} {426}},\ \bibinfo {pages} {537} (\bibinfo {year}
  {2003})}\BibitemShut {NoStop}%
\bibitem [{\citenamefont {Mewes}\ \emph {et~al.}(1997)\citenamefont {Mewes},
  \citenamefont {Andrews}, \citenamefont {Kurn}, \citenamefont {Durfee},
  \citenamefont {Townsend},\ and\ \citenamefont {Ketterle}}]{Mewes:1997}%
  \BibitemOpen
  \bibfield  {author} {\bibinfo {author} {\bibfnamefont {M.-O.}\ \bibnamefont
  {Mewes}}, \bibinfo {author} {\bibfnamefont {M.~R.}\ \bibnamefont {Andrews}},
  \bibinfo {author} {\bibfnamefont {D.~M.}\ \bibnamefont {Kurn}}, \bibinfo
  {author} {\bibfnamefont {D.~S.}\ \bibnamefont {Durfee}}, \bibinfo {author}
  {\bibfnamefont {C.~G.}\ \bibnamefont {Townsend}}, \ and\ \bibinfo {author}
  {\bibfnamefont {W.}~\bibnamefont {Ketterle}},\ }\href@noop {} {\bibfield
  {journal} {\bibinfo  {journal} {Physical Review Letters}\ }\textbf {\bibinfo
  {volume} {78}},\ \bibinfo {pages} {582} (\bibinfo {year} {1997})}\BibitemShut
  {NoStop}%
\bibitem [{\citenamefont {Deng}\ \emph {et~al.}(1999)\citenamefont {Deng},
  \citenamefont {Hagley}, \citenamefont {Wen}, \citenamefont {Trippenbach},
  \citenamefont {Band}, \citenamefont {Julienne}, \citenamefont {Simsarian},
  \citenamefont {Helmerson}, \citenamefont {Rolston},\ and\ \citenamefont
  {Phillips}}]{Deng:1999}%
  \BibitemOpen
  \bibfield  {author} {\bibinfo {author} {\bibfnamefont {L.}~\bibnamefont
  {Deng}}, \bibinfo {author} {\bibfnamefont {E.~W.}\ \bibnamefont {Hagley}},
  \bibinfo {author} {\bibfnamefont {J.}~\bibnamefont {Wen}}, \bibinfo {author}
  {\bibfnamefont {M.}~\bibnamefont {Trippenbach}}, \bibinfo {author}
  {\bibfnamefont {Y.}~\bibnamefont {Band}}, \bibinfo {author} {\bibfnamefont
  {P.~S.}\ \bibnamefont {Julienne}}, \bibinfo {author} {\bibfnamefont {J.~E.}\
  \bibnamefont {Simsarian}}, \bibinfo {author} {\bibfnamefont {K.}~\bibnamefont
  {Helmerson}}, \bibinfo {author} {\bibfnamefont {S.~L.}\ \bibnamefont
  {Rolston}}, \ and\ \bibinfo {author} {\bibfnamefont {W.~D.}\ \bibnamefont
  {Phillips}},\ }\href@noop {} {\bibfield  {journal} {\bibinfo  {journal}
  {Nature}\ }\textbf {\bibinfo {volume} {398}},\ \bibinfo {pages} {218}
  (\bibinfo {year} {1999})}\BibitemShut {NoStop}%
\bibitem [{\citenamefont {Perrin}\ \emph {et~al.}(2007)\citenamefont {Perrin},
  \citenamefont {Chang}, \citenamefont {Krachmalnicoff}, \citenamefont
  {Schellekens}, \citenamefont {Boiron}, \citenamefont {Aspect},\ and\
  \citenamefont {Westbrook}}]{Perrin:2007}%
  \BibitemOpen
  \bibfield  {author} {\bibinfo {author} {\bibfnamefont {A.}~\bibnamefont
  {Perrin}}, \bibinfo {author} {\bibfnamefont {H.}~\bibnamefont {Chang}},
  \bibinfo {author} {\bibfnamefont {V.}~\bibnamefont {Krachmalnicoff}},
  \bibinfo {author} {\bibfnamefont {M.}~\bibnamefont {Schellekens}}, \bibinfo
  {author} {\bibfnamefont {D.}~\bibnamefont {Boiron}}, \bibinfo {author}
  {\bibfnamefont {A.}~\bibnamefont {Aspect}}, \ and\ \bibinfo {author}
  {\bibfnamefont {C.~I.}\ \bibnamefont {Westbrook}},\ }\href@noop {} {\bibfield
   {journal} {\bibinfo  {journal} {Physical Review Letters}\ }\textbf {\bibinfo
  {volume} {99}},\ \bibinfo {pages} {150405} (\bibinfo {year}
  {2007})}\BibitemShut {NoStop}%
\bibitem [{\citenamefont {Gericke}\ \emph {et~al.}(2008)\citenamefont
  {Gericke}, \citenamefont {W{\"u}rtz}, \citenamefont {Reitz}, \citenamefont
  {Langen},\ and\ \citenamefont {Ott}}]{Gericke:2008}%
  \BibitemOpen
  \bibfield  {author} {\bibinfo {author} {\bibfnamefont {T.}~\bibnamefont
  {Gericke}}, \bibinfo {author} {\bibfnamefont {P.}~\bibnamefont {W{\"u}rtz}},
  \bibinfo {author} {\bibfnamefont {D.}~\bibnamefont {Reitz}}, \bibinfo
  {author} {\bibfnamefont {T.}~\bibnamefont {Langen}}, \ and\ \bibinfo {author}
  {\bibfnamefont {H.}~\bibnamefont {Ott}},\ }\href@noop {} {\bibfield
  {journal} {\bibinfo  {journal} {Nature Physics}\ }\textbf {\bibinfo {volume}
  {4}},\ \bibinfo {pages} {949} (\bibinfo {year} {2008})}\BibitemShut {NoStop}%
\bibitem [{\citenamefont {Bakr}\ \emph {et~al.}(2009)\citenamefont {Bakr},
  \citenamefont {Gillen}, \citenamefont {Peng}, \citenamefont {F{\"o}lling},\
  and\ \citenamefont {Greiner}}]{Bakr:2009}%
  \BibitemOpen
  \bibfield  {author} {\bibinfo {author} {\bibfnamefont {W.~S.}\ \bibnamefont
  {Bakr}}, \bibinfo {author} {\bibfnamefont {J.~I.}\ \bibnamefont {Gillen}},
  \bibinfo {author} {\bibfnamefont {A.}~\bibnamefont {Peng}}, \bibinfo {author}
  {\bibfnamefont {S.}~\bibnamefont {F{\"o}lling}}, \ and\ \bibinfo {author}
  {\bibfnamefont {M.}~\bibnamefont {Greiner}},\ }\href@noop {} {\bibfield
  {journal} {\bibinfo  {journal} {Nature}\ }\textbf {\bibinfo {volume} {462}},\
  \bibinfo {pages} {74} (\bibinfo {year} {2009})}\BibitemShut {NoStop}%
\bibitem [{\citenamefont {Sherson}\ \emph {et~al.}(2010)\citenamefont
  {Sherson}, \citenamefont {Weitenberg}, \citenamefont {Endres}, \citenamefont
  {Cheneau}, \citenamefont {Bloch},\ and\ \citenamefont {Kuhr}}]{Sherson:2010}%
  \BibitemOpen
  \bibfield  {author} {\bibinfo {author} {\bibfnamefont {J.~F.}\ \bibnamefont
  {Sherson}}, \bibinfo {author} {\bibfnamefont {C.}~\bibnamefont {Weitenberg}},
  \bibinfo {author} {\bibfnamefont {M.}~\bibnamefont {Endres}}, \bibinfo
  {author} {\bibfnamefont {M.}~\bibnamefont {Cheneau}}, \bibinfo {author}
  {\bibfnamefont {I.}~\bibnamefont {Bloch}}, \ and\ \bibinfo {author}
  {\bibfnamefont {S.}~\bibnamefont {Kuhr}},\ }\href@noop {} {\bibfield
  {journal} {\bibinfo  {journal} {Nature}\ }\textbf {\bibinfo {volume} {467}},\
  \bibinfo {pages} {68} (\bibinfo {year} {2010})}\BibitemShut {NoStop}%
\bibitem [{\citenamefont {Schellekens}\ \emph {et~al.}(2005)\citenamefont
  {Schellekens}, \citenamefont {Hoppeler}, \citenamefont {Perrin},
  \citenamefont {Viana~Gomes}, \citenamefont {Boiron}, \citenamefont {Aspect},\
  and\ \citenamefont {Westbrook}}]{Schellekens:2005}%
  \BibitemOpen
  \bibfield  {author} {\bibinfo {author} {\bibfnamefont {M.}~\bibnamefont
  {Schellekens}}, \bibinfo {author} {\bibfnamefont {R.}~\bibnamefont
  {Hoppeler}}, \bibinfo {author} {\bibfnamefont {A.}~\bibnamefont {Perrin}},
  \bibinfo {author} {\bibfnamefont {J.}~\bibnamefont {Viana~Gomes}}, \bibinfo
  {author} {\bibfnamefont {D.}~\bibnamefont {Boiron}}, \bibinfo {author}
  {\bibfnamefont {A.}~\bibnamefont {Aspect}}, \ and\ \bibinfo {author}
  {\bibfnamefont {C.~I.}\ \bibnamefont {Westbrook}},\ }\href@noop {} {\bibfield
   {journal} {\bibinfo  {journal} {Science}\ }\textbf {\bibinfo {volume}
  {310}},\ \bibinfo {pages} {648} (\bibinfo {year} {2005})}\BibitemShut
  {NoStop}%
\bibitem [{\citenamefont {B{\"u}cker}\ \emph {et~al.}(2009)\citenamefont
  {B{\"u}cker}, \citenamefont {Perrin}, \citenamefont {Manz}, \citenamefont
  {Betz}, \citenamefont {Koller}, \citenamefont {Plisson}, \citenamefont
  {Rottmann}, \citenamefont {Schumm},\ and\ \citenamefont
  {Schmiedmayer}}]{Buecker:2009}%
  \BibitemOpen
  \bibfield  {author} {\bibinfo {author} {\bibfnamefont {R.}~\bibnamefont
  {B{\"u}cker}}, \bibinfo {author} {\bibfnamefont {A.}~\bibnamefont {Perrin}},
  \bibinfo {author} {\bibfnamefont {S.}~\bibnamefont {Manz}}, \bibinfo {author}
  {\bibfnamefont {T.}~\bibnamefont {Betz}}, \bibinfo {author} {\bibfnamefont
  {C.}~\bibnamefont {Koller}}, \bibinfo {author} {\bibfnamefont
  {T.}~\bibnamefont {Plisson}}, \bibinfo {author} {\bibfnamefont
  {J.}~\bibnamefont {Rottmann}}, \bibinfo {author} {\bibfnamefont
  {T.}~\bibnamefont {Schumm}}, \ and\ \bibinfo {author} {\bibfnamefont
  {J.}~\bibnamefont {Schmiedmayer}},\ }\href@noop {} {\bibfield  {journal}
  {\bibinfo  {journal} {New Journal of Physics}\ }\textbf {\bibinfo {volume}
  {11}},\ \bibinfo {pages} {103039} (\bibinfo {year} {2009})}\BibitemShut
  {NoStop}%
\bibitem [{\citenamefont {Robert}\ \emph {et~al.}(2001)\citenamefont {Robert},
  \citenamefont {Sirjean}, \citenamefont {Browaeys}, \citenamefont {Poupard},
  \citenamefont {Nowak}, \citenamefont {Boiron}, \citenamefont {Westbrook},\
  and\ \citenamefont {Aspect}}]{Robert:2001}%
  \BibitemOpen
  \bibfield  {author} {\bibinfo {author} {\bibfnamefont {A.}~\bibnamefont
  {Robert}}, \bibinfo {author} {\bibfnamefont {O.}~\bibnamefont {Sirjean}},
  \bibinfo {author} {\bibfnamefont {A.}~\bibnamefont {Browaeys}}, \bibinfo
  {author} {\bibfnamefont {J.}~\bibnamefont {Poupard}}, \bibinfo {author}
  {\bibfnamefont {S.}~\bibnamefont {Nowak}}, \bibinfo {author} {\bibfnamefont
  {D.}~\bibnamefont {Boiron}}, \bibinfo {author} {\bibfnamefont {C.~I.}\
  \bibnamefont {Westbrook}}, \ and\ \bibinfo {author} {\bibfnamefont
  {A.}~\bibnamefont {Aspect}},\ }\href@noop {} {\bibfield  {journal} {\bibinfo
  {journal} {Science}\ }\textbf {\bibinfo {volume} {292}},\ \bibinfo {pages}
  {461} (\bibinfo {year} {2001})}\BibitemShut {NoStop}%
\bibitem [{\citenamefont {Pereira Dos~Santos}\ \emph
  {et~al.}(2001)\citenamefont {Pereira Dos~Santos}, \citenamefont
  {L\'{e}onard}, \citenamefont {Wang}, \citenamefont {Barrelet}, \citenamefont
  {Perales}, \citenamefont {Rasel}, \citenamefont {Unnikrishnan}, \citenamefont
  {Leduc},\ and\ \citenamefont {Cohen-Tannoudji}}]{Santos:2001}%
  \BibitemOpen
  \bibfield  {author} {\bibinfo {author} {\bibfnamefont {F.}~\bibnamefont
  {Pereira Dos~Santos}}, \bibinfo {author} {\bibfnamefont {J.}~\bibnamefont
  {L\'{e}onard}}, \bibinfo {author} {\bibfnamefont {J.}~\bibnamefont {Wang}},
  \bibinfo {author} {\bibfnamefont {C.~J.}\ \bibnamefont {Barrelet}}, \bibinfo
  {author} {\bibfnamefont {F.}~\bibnamefont {Perales}}, \bibinfo {author}
  {\bibfnamefont {E.}~\bibnamefont {Rasel}}, \bibinfo {author} {\bibfnamefont
  {C.~S.}\ \bibnamefont {Unnikrishnan}}, \bibinfo {author} {\bibfnamefont
  {M.}~\bibnamefont {Leduc}}, \ and\ \bibinfo {author} {\bibfnamefont
  {C.}~\bibnamefont {Cohen-Tannoudji}},\ }\href@noop {} {\bibfield  {journal}
  {\bibinfo  {journal} {Physical Review Letters}\ }\textbf {\bibinfo {volume}
  {86}},\ \bibinfo {pages} {3459} (\bibinfo {year} {2001})}\BibitemShut
  {NoStop}%
\bibitem [{\citenamefont {Vassen}\ \emph {et~al.}(2012)\citenamefont {Vassen},
  \citenamefont {Cohen-Tannoudji}, \citenamefont {Leduc}, \citenamefont
  {Boiron}, \citenamefont {Westbrook}, \citenamefont {Truscott}, \citenamefont
  {Baldwin}, \citenamefont {Birkl}, \citenamefont {Cancio},\ and\ \citenamefont
  {Trippenbach}}]{Vassen:2012}%
  \BibitemOpen
  \bibfield  {author} {\bibinfo {author} {\bibfnamefont {W.}~\bibnamefont
  {Vassen}}, \bibinfo {author} {\bibfnamefont {C.}~\bibnamefont
  {Cohen-Tannoudji}}, \bibinfo {author} {\bibfnamefont {M.}~\bibnamefont
  {Leduc}}, \bibinfo {author} {\bibfnamefont {D.}~\bibnamefont {Boiron}},
  \bibinfo {author} {\bibfnamefont {C.~I.}\ \bibnamefont {Westbrook}}, \bibinfo
  {author} {\bibfnamefont {A.}~\bibnamefont {Truscott}}, \bibinfo {author}
  {\bibfnamefont {K.}~\bibnamefont {Baldwin}}, \bibinfo {author} {\bibfnamefont
  {G.}~\bibnamefont {Birkl}}, \bibinfo {author} {\bibfnamefont
  {P.}~\bibnamefont {Cancio}}, \ and\ \bibinfo {author} {\bibfnamefont
  {M.}~\bibnamefont {Trippenbach}},\ }\href@noop {} {\bibfield  {journal}
  {\bibinfo  {journal} {Reviews of Modern Physics}\ }\textbf {\bibinfo {volume}
  {84}},\ \bibinfo {pages} {175} (\bibinfo {year} {2012})}\BibitemShut
  {NoStop}%
\bibitem [{\citenamefont {Jeltes}\ \emph {et~al.}(2007)\citenamefont {Jeltes},
  \citenamefont {McNamara}, \citenamefont {Hogervorst}, \citenamefont {Vassen},
  \citenamefont {Krachmalnicoff}, \citenamefont {Schellekens}, \citenamefont
  {Perrin}, \citenamefont {Chang}, \citenamefont {Boiron}, \citenamefont
  {Aspect},\ and\ \citenamefont {Westbrook}}]{Jeltes:2007}%
  \BibitemOpen
  \bibfield  {author} {\bibinfo {author} {\bibfnamefont {T.}~\bibnamefont
  {Jeltes}}, \bibinfo {author} {\bibfnamefont {J.~M.}\ \bibnamefont
  {McNamara}}, \bibinfo {author} {\bibfnamefont {W.}~\bibnamefont
  {Hogervorst}}, \bibinfo {author} {\bibfnamefont {W.}~\bibnamefont {Vassen}},
  \bibinfo {author} {\bibfnamefont {V.}~\bibnamefont {Krachmalnicoff}},
  \bibinfo {author} {\bibfnamefont {M.}~\bibnamefont {Schellekens}}, \bibinfo
  {author} {\bibfnamefont {A.}~\bibnamefont {Perrin}}, \bibinfo {author}
  {\bibfnamefont {H.}~\bibnamefont {Chang}}, \bibinfo {author} {\bibfnamefont
  {D.}~\bibnamefont {Boiron}}, \bibinfo {author} {\bibfnamefont
  {A.}~\bibnamefont {Aspect}}, \ and\ \bibinfo {author} {\bibfnamefont {C.~I.}\
  \bibnamefont {Westbrook}},\ }\href@noop {} {\bibfield  {journal} {\bibinfo
  {journal} {Nature}\ }\textbf {\bibinfo {volume} {445}},\ \bibinfo {pages}
  {402} (\bibinfo {year} {2007})}\BibitemShut {NoStop}%
\bibitem [{\citenamefont {Hodgman}\ \emph {et~al.}(2011)\citenamefont
  {Hodgman}, \citenamefont {Dall}, \citenamefont {Manning}, \citenamefont
  {Baldwin},\ and\ \citenamefont {Truscott}}]{Hodgman:2011}%
  \BibitemOpen
  \bibfield  {author} {\bibinfo {author} {\bibfnamefont {S.~S.}\ \bibnamefont
  {Hodgman}}, \bibinfo {author} {\bibfnamefont {R.~G.}\ \bibnamefont {Dall}},
  \bibinfo {author} {\bibfnamefont {A.~G.}\ \bibnamefont {Manning}}, \bibinfo
  {author} {\bibfnamefont {K.~G.~H.}\ \bibnamefont {Baldwin}}, \ and\ \bibinfo
  {author} {\bibfnamefont {A.~G.}\ \bibnamefont {Truscott}},\ }\href@noop {}
  {\bibfield  {journal} {\bibinfo  {journal} {Science}\ }\textbf {\bibinfo
  {volume} {331}},\ \bibinfo {pages} {1046} (\bibinfo {year}
  {2011})}\BibitemShut {NoStop}%
\bibitem [{\citenamefont {Ferris}\ \emph {et~al.}(2008)\citenamefont {Ferris},
  \citenamefont {Olsen}, \citenamefont {Cavalcanti},\ and\ \citenamefont
  {Davis}}]{Ferris:2008}%
  \BibitemOpen
  \bibfield  {author} {\bibinfo {author} {\bibfnamefont {A.~J.}\ \bibnamefont
  {Ferris}}, \bibinfo {author} {\bibfnamefont {M.~K.}\ \bibnamefont {Olsen}},
  \bibinfo {author} {\bibfnamefont {E.~G.}\ \bibnamefont {Cavalcanti}}, \ and\
  \bibinfo {author} {\bibfnamefont {M.~J.}\ \bibnamefont {Davis}},\ }\href@noop
  {} {\bibfield  {journal} {\bibinfo  {journal} {Physical Review A}\ }\textbf
  {\bibinfo {volume} {78}},\ \bibinfo {pages} {060104} (\bibinfo {year}
  {2008})}\BibitemShut {NoStop}%
\bibitem [{\citenamefont {Ferris}\ \emph {et~al.}(2009)\citenamefont {Ferris},
  \citenamefont {Olsen},\ and\ \citenamefont {Davis}}]{Ferris:2009}%
  \BibitemOpen
  \bibfield  {author} {\bibinfo {author} {\bibfnamefont {A.~J.}\ \bibnamefont
  {Ferris}}, \bibinfo {author} {\bibfnamefont {M.~K.}\ \bibnamefont {Olsen}}, \
  and\ \bibinfo {author} {\bibfnamefont {M.~J.}\ \bibnamefont {Davis}},\
  }\href@noop {} {\bibfield  {journal} {\bibinfo  {journal} {Physical Review
  A}\ }\textbf {\bibinfo {volume} {79}},\ \bibinfo {pages} {043634} (\bibinfo
  {year} {2009})}\BibitemShut {NoStop}%
\bibitem [{\citenamefont {Kofler}\ \emph {et~al.}(2012)\citenamefont {Kofler},
  \citenamefont {Singh}, \citenamefont {Ebner}, \citenamefont {Keller},
  \citenamefont {Kotyrba},\ and\ \citenamefont {Zeilinger}}]{Kofler:2012}%
  \BibitemOpen
  \bibfield  {author} {\bibinfo {author} {\bibfnamefont {J.}~\bibnamefont
  {Kofler}}, \bibinfo {author} {\bibfnamefont {M.}~\bibnamefont {Singh}},
  \bibinfo {author} {\bibfnamefont {M.}~\bibnamefont {Ebner}}, \bibinfo
  {author} {\bibfnamefont {M.}~\bibnamefont {Keller}}, \bibinfo {author}
  {\bibfnamefont {M.}~\bibnamefont {Kotyrba}}, \ and\ \bibinfo {author}
  {\bibfnamefont {A.}~\bibnamefont {Zeilinger}},\ }\href@noop {} {\bibfield
  {journal} {\bibinfo  {journal} {Physical Review A}\ }\textbf {\bibinfo
  {volume} {86}},\ \bibinfo {pages} {032115} (\bibinfo {year}
  {2012})}\BibitemShut {NoStop}%
\bibitem [{\citenamefont {Rarity}\ and\ \citenamefont
  {Tapster}(1990)}]{Rarity:1990}%
  \BibitemOpen
  \bibfield  {author} {\bibinfo {author} {\bibfnamefont {J.~G.}\ \bibnamefont
  {Rarity}}\ and\ \bibinfo {author} {\bibfnamefont {P.~R.}\ \bibnamefont
  {Tapster}},\ }\href@noop {} {\bibfield  {journal} {\bibinfo  {journal}
  {Physical Review Letters}\ }\textbf {\bibinfo {volume} {64}},\ \bibinfo
  {pages} {2495} (\bibinfo {year} {1990})}\BibitemShut {NoStop}%
\bibitem [{\citenamefont {Ou}\ \emph {et~al.}(1990)\citenamefont {Ou},
  \citenamefont {Zou}, \citenamefont {Wang},\ and\ \citenamefont
  {Mandel}}]{Ou:1990}%
  \BibitemOpen
  \bibfield  {author} {\bibinfo {author} {\bibfnamefont {Z.~Y.}\ \bibnamefont
  {Ou}}, \bibinfo {author} {\bibfnamefont {X.~Y.}\ \bibnamefont {Zou}},
  \bibinfo {author} {\bibfnamefont {L.~J.}\ \bibnamefont {Wang}}, \ and\
  \bibinfo {author} {\bibfnamefont {L.}~\bibnamefont {Mandel}},\ }\href@noop {}
  {\bibfield  {journal} {\bibinfo  {journal} {Physical Review Letters}\
  }\textbf {\bibinfo {volume} {65}},\ \bibinfo {pages} {321} (\bibinfo {year}
  {1990})}\BibitemShut {NoStop}%
\bibitem [{\citenamefont {Ou}\ \emph {et~al.}(1992)\citenamefont {Ou},
  \citenamefont {Pereira}, \citenamefont {Kimble},\ and\ \citenamefont
  {Peng}}]{Ou:1992}%
  \BibitemOpen
  \bibfield  {author} {\bibinfo {author} {\bibfnamefont {Z.~Y.}\ \bibnamefont
  {Ou}}, \bibinfo {author} {\bibfnamefont {S.~F.}\ \bibnamefont {Pereira}},
  \bibinfo {author} {\bibfnamefont {H.~J.}\ \bibnamefont {Kimble}}, \ and\
  \bibinfo {author} {\bibfnamefont {K.~C.}\ \bibnamefont {Peng}},\ }\href@noop
  {} {\bibfield  {journal} {\bibinfo  {journal} {Physical Review Letters}\
  }\textbf {\bibinfo {volume} {68}},\ \bibinfo {pages} {3663} (\bibinfo {year}
  {1992})}\BibitemShut {NoStop}%
\bibitem [{\citenamefont {Vogels}\ \emph {et~al.}(2002)\citenamefont {Vogels},
  \citenamefont {Xu},\ and\ \citenamefont {Ketterle}}]{Vogels:2002}%
  \BibitemOpen
  \bibfield  {author} {\bibinfo {author} {\bibfnamefont {J.~M.}\ \bibnamefont
  {Vogels}}, \bibinfo {author} {\bibfnamefont {K.}~\bibnamefont {Xu}}, \ and\
  \bibinfo {author} {\bibfnamefont {W.}~\bibnamefont {Ketterle}},\ }\href@noop
  {} {\bibfield  {journal} {\bibinfo  {journal} {Physical Review Letters}\
  }\textbf {\bibinfo {volume} {89}},\ \bibinfo {pages} {020401} (\bibinfo
  {year} {2002})}\BibitemShut {NoStop}%
\bibitem [{\citenamefont {Vogels}\ \emph {et~al.}(2003)\citenamefont {Vogels},
  \citenamefont {Chin},\ and\ \citenamefont {Ketterle}}]{Vogels:2003}%
  \BibitemOpen
  \bibfield  {author} {\bibinfo {author} {\bibfnamefont {J.~M.}\ \bibnamefont
  {Vogels}}, \bibinfo {author} {\bibfnamefont {J.~K.}\ \bibnamefont {Chin}}, \
  and\ \bibinfo {author} {\bibfnamefont {W.}~\bibnamefont {Ketterle}},\
  }\href@noop {} {\bibfield  {journal} {\bibinfo  {journal} {Physical Review
  Letters}\ }\textbf {\bibinfo {volume} {90}},\ \bibinfo {pages} {030403}
  (\bibinfo {year} {2003})}\BibitemShut {NoStop}%
\bibitem [{\citenamefont {Kheruntsyan}\ \emph {et~al.}(2012)\citenamefont
  {Kheruntsyan}, \citenamefont {Jaskula}, \citenamefont {Deuar}, \citenamefont
  {Bonneau}, \citenamefont {Partridge}, \citenamefont {Ruaudel}, \citenamefont
  {Lopes}, \citenamefont {Boiron},\ and\ \citenamefont
  {Westbrook}}]{Kheruntsyan:2012}%
  \BibitemOpen
  \bibfield  {author} {\bibinfo {author} {\bibfnamefont {K.~V.}\ \bibnamefont
  {Kheruntsyan}}, \bibinfo {author} {\bibfnamefont {J.-C.}\ \bibnamefont
  {Jaskula}}, \bibinfo {author} {\bibfnamefont {P.}~\bibnamefont {Deuar}},
  \bibinfo {author} {\bibfnamefont {M.}~\bibnamefont {Bonneau}}, \bibinfo
  {author} {\bibfnamefont {G.~B.}\ \bibnamefont {Partridge}}, \bibinfo {author}
  {\bibfnamefont {J.}~\bibnamefont {Ruaudel}}, \bibinfo {author} {\bibfnamefont
  {R.}~\bibnamefont {Lopes}}, \bibinfo {author} {\bibfnamefont
  {D.}~\bibnamefont {Boiron}}, \ and\ \bibinfo {author} {\bibfnamefont {C.~I.}\
  \bibnamefont {Westbrook}},\ }\href@noop {} {\bibfield  {journal} {\bibinfo
  {journal} {Physical Review Letters}\ }\textbf {\bibinfo {volume} {108}},\
  \bibinfo {pages} {260401} (\bibinfo {year} {2012})}\BibitemShut {NoStop}%
\bibitem [{\citenamefont {Einstein}\ \emph {et~al.}(1935)\citenamefont
  {Einstein}, \citenamefont {Podolsky},\ and\ \citenamefont
  {Rosen}}]{Einstein:1935}%
  \BibitemOpen
  \bibfield  {author} {\bibinfo {author} {\bibfnamefont {A.}~\bibnamefont
  {Einstein}}, \bibinfo {author} {\bibfnamefont {B.}~\bibnamefont {Podolsky}},
  \ and\ \bibinfo {author} {\bibfnamefont {N.}~\bibnamefont {Rosen}},\
  }\href@noop {} {\bibfield  {journal} {\bibinfo  {journal} {Physical Review}\
  }\textbf {\bibinfo {volume} {47}},\ \bibinfo {pages} {777} (\bibinfo {year}
  {1935})}\BibitemShut {NoStop}%
\bibitem [{\citenamefont {Dall}\ \emph {et~al.}(2009)\citenamefont {Dall},
  \citenamefont {Byron}, \citenamefont {Truscott}, \citenamefont {Dennis},
  \citenamefont {Johnsson},\ and\ \citenamefont {Hope}}]{Dall:2009}%
  \BibitemOpen
  \bibfield  {author} {\bibinfo {author} {\bibfnamefont {R.~G.}\ \bibnamefont
  {Dall}}, \bibinfo {author} {\bibfnamefont {L.~J.}\ \bibnamefont {Byron}},
  \bibinfo {author} {\bibfnamefont {A.~G.}\ \bibnamefont {Truscott}}, \bibinfo
  {author} {\bibfnamefont {G.~R.}\ \bibnamefont {Dennis}}, \bibinfo {author}
  {\bibfnamefont {M.~T.}\ \bibnamefont {Johnsson}}, \ and\ \bibinfo {author}
  {\bibfnamefont {J.~J.}\ \bibnamefont {Hope}},\ }\href@noop {} {\bibfield
  {journal} {\bibinfo  {journal} {Physical Review A}\ }\textbf {\bibinfo
  {volume} {79}},\ \bibinfo {pages} {011601(R)} (\bibinfo {year}
  {2009})}\BibitemShut {NoStop}%
\bibitem [{\citenamefont {B{\"u}cker}\ \emph {et~al.}(2011)\citenamefont
  {B{\"u}cker}, \citenamefont {Grond}, \citenamefont {Manz}, \citenamefont
  {Berrada}, \citenamefont {Betz}, \citenamefont {Koller}, \citenamefont
  {Hohenester}, \citenamefont {Schumm}, \citenamefont {Perrin},\ and\
  \citenamefont {Schmiedmayer}}]{Buecker:2011}%
  \BibitemOpen
  \bibfield  {author} {\bibinfo {author} {\bibfnamefont {R.}~\bibnamefont
  {B{\"u}cker}}, \bibinfo {author} {\bibfnamefont {J.}~\bibnamefont {Grond}},
  \bibinfo {author} {\bibfnamefont {S.}~\bibnamefont {Manz}}, \bibinfo {author}
  {\bibfnamefont {T.}~\bibnamefont {Berrada}}, \bibinfo {author} {\bibfnamefont
  {T.}~\bibnamefont {Betz}}, \bibinfo {author} {\bibfnamefont {C.}~\bibnamefont
  {Koller}}, \bibinfo {author} {\bibfnamefont {U.}~\bibnamefont {Hohenester}},
  \bibinfo {author} {\bibfnamefont {T.}~\bibnamefont {Schumm}}, \bibinfo
  {author} {\bibfnamefont {A.}~\bibnamefont {Perrin}}, \ and\ \bibinfo {author}
  {\bibfnamefont {J.}~\bibnamefont {Schmiedmayer}},\ }\href@noop {} {\bibfield
  {journal} {\bibinfo  {journal} {Nature Physics}\ }\textbf {\bibinfo {volume}
  {7}},\ \bibinfo {pages} {608} (\bibinfo {year} {2011})}\BibitemShut {NoStop}%
\bibitem [{\citenamefont {Drake}(1971)}]{Drake:1971}%
  \BibitemOpen
  \bibfield  {author} {\bibinfo {author} {\bibfnamefont {G.~W.~F.}\
  \bibnamefont {Drake}},\ }\href@noop {} {\bibfield  {journal} {\bibinfo
  {journal} {Physical Review A}\ }\textbf {\bibinfo {volume} {3}},\ \bibinfo
  {pages} {908} (\bibinfo {year} {1971})}\BibitemShut {NoStop}%
\bibitem [{\citenamefont {Moos}\ and\ \citenamefont
  {Woodworth}(1973)}]{Moos:1973}%
  \BibitemOpen
  \bibfield  {author} {\bibinfo {author} {\bibfnamefont {H.~W.}\ \bibnamefont
  {Moos}}\ and\ \bibinfo {author} {\bibfnamefont {J.~R.}\ \bibnamefont
  {Woodworth}},\ }\href@noop {} {\bibfield  {journal} {\bibinfo  {journal}
  {Physical Review Letters}\ }\textbf {\bibinfo {volume} {30}},\ \bibinfo
  {pages} {775} (\bibinfo {year} {1973})}\BibitemShut {NoStop}%
\bibitem [{\citenamefont {Hodgman}\ \emph {et~al.}(2009)\citenamefont
  {Hodgman}, \citenamefont {Dall}, \citenamefont {Byron}, \citenamefont
  {Baldwin}, \citenamefont {Buckman},\ and\ \citenamefont
  {Truscott}}]{Hodgman:2009}%
  \BibitemOpen
  \bibfield  {author} {\bibinfo {author} {\bibfnamefont {S.~S.}\ \bibnamefont
  {Hodgman}}, \bibinfo {author} {\bibfnamefont {R.~G.}\ \bibnamefont {Dall}},
  \bibinfo {author} {\bibfnamefont {L.~J.}\ \bibnamefont {Byron}}, \bibinfo
  {author} {\bibfnamefont {K.~G.~H.}\ \bibnamefont {Baldwin}}, \bibinfo
  {author} {\bibfnamefont {S.~J.}\ \bibnamefont {Buckman}}, \ and\ \bibinfo
  {author} {\bibfnamefont {A.~G.}\ \bibnamefont {Truscott}},\ }\href@noop {}
  {\bibfield  {journal} {\bibinfo  {journal} {Physical Review Letters}\
  }\textbf {\bibinfo {volume} {103}},\ \bibinfo {pages} {053002} (\bibinfo
  {year} {2009})}\BibitemShut {NoStop}%
\bibitem [{\citenamefont {Kawanaka}\ \emph {et~al.}(1993)\citenamefont
  {Kawanaka}, \citenamefont {Hagiuda}, \citenamefont {Shimizu}, \citenamefont
  {Shimizu},\ and\ \citenamefont {Takuma}}]{Kawanaka:1993}%
  \BibitemOpen
  \bibfield  {author} {\bibinfo {author} {\bibfnamefont {J.}~\bibnamefont
  {Kawanaka}}, \bibinfo {author} {\bibfnamefont {M.}~\bibnamefont {Hagiuda}},
  \bibinfo {author} {\bibfnamefont {K.}~\bibnamefont {Shimizu}}, \bibinfo
  {author} {\bibfnamefont {F.}~\bibnamefont {Shimizu}}, \ and\ \bibinfo
  {author} {\bibfnamefont {H.}~\bibnamefont {Takuma}},\ }\href@noop {}
  {\bibfield  {journal} {\bibinfo  {journal} {Applied Physics B}\ }\textbf
  {\bibinfo {volume} {56}},\ \bibinfo {pages} {21} (\bibinfo {year}
  {1993})}\BibitemShut {NoStop}%
\bibitem [{\citenamefont {Aspect}\ \emph {et~al.}(1990)\citenamefont {Aspect},
  \citenamefont {Vansteenkiste}, \citenamefont {Kaiser}, \citenamefont
  {Haberland},\ and\ \citenamefont {Karrais}}]{Aspect:1990}%
  \BibitemOpen
  \bibfield  {author} {\bibinfo {author} {\bibfnamefont {A.}~\bibnamefont
  {Aspect}}, \bibinfo {author} {\bibfnamefont {N.}~\bibnamefont
  {Vansteenkiste}}, \bibinfo {author} {\bibfnamefont {R.}~\bibnamefont
  {Kaiser}}, \bibinfo {author} {\bibfnamefont {H.}~\bibnamefont {Haberland}}, \
  and\ \bibinfo {author} {\bibfnamefont {M.}~\bibnamefont {Karrais}},\
  }\href@noop {} {\bibfield  {journal} {\bibinfo  {journal} {Chemical Physics}\
  }\textbf {\bibinfo {volume} {145}},\ \bibinfo {pages} {307} (\bibinfo {year}
  {1990})}\BibitemShut {NoStop}%
\bibitem [{\citenamefont {Phillips}\ and\ \citenamefont
  {Metcalf}(1982)}]{Phillips:1982}%
  \BibitemOpen
  \bibfield  {author} {\bibinfo {author} {\bibfnamefont {W.~D.}\ \bibnamefont
  {Phillips}}\ and\ \bibinfo {author} {\bibfnamefont {H.}~\bibnamefont
  {Metcalf}},\ }\href@noop {} {\bibfield  {journal} {\bibinfo  {journal}
  {Physical Review Letters}\ }\textbf {\bibinfo {volume} {48}},\ \bibinfo
  {pages} {596} (\bibinfo {year} {1982})}\BibitemShut {NoStop}%
\bibitem [{\citenamefont {Dedman}\ \emph {et~al.}(2004)\citenamefont {Dedman},
  \citenamefont {Nes}, \citenamefont {Hanna}, \citenamefont {Dall},
  \citenamefont {Baldwin},\ and\ \citenamefont {Truscott}}]{Dedman:2004}%
  \BibitemOpen
  \bibfield  {author} {\bibinfo {author} {\bibfnamefont {C.~J.}\ \bibnamefont
  {Dedman}}, \bibinfo {author} {\bibfnamefont {J.}~\bibnamefont {Nes}},
  \bibinfo {author} {\bibfnamefont {T.~M.}\ \bibnamefont {Hanna}}, \bibinfo
  {author} {\bibfnamefont {R.~G.}\ \bibnamefont {Dall}}, \bibinfo {author}
  {\bibfnamefont {K.~G.~H.}\ \bibnamefont {Baldwin}}, \ and\ \bibinfo {author}
  {\bibfnamefont {A.~G.}\ \bibnamefont {Truscott}},\ }\href@noop {} {\bibfield
  {journal} {\bibinfo  {journal} {Review of Scientific Instruments}\ }\textbf
  {\bibinfo {volume} {75}},\ \bibinfo {pages} {5136} (\bibinfo {year}
  {2004})}\BibitemShut {NoStop}%
\bibitem [{\citenamefont {Penning}(1927)}]{Penning:1927}%
  \BibitemOpen
  \bibfield  {author} {\bibinfo {author} {\bibfnamefont {F.~M.}\ \bibnamefont
  {Penning}},\ }\href@noop {} {\bibfield  {journal} {\bibinfo  {journal} {Die
  Naturwissenschaften}\ }\textbf {\bibinfo {volume} {15}},\ \bibinfo {pages}
  {818} (\bibinfo {year} {1927})}\BibitemShut {NoStop}%
\bibitem [{\citenamefont {Pritchard}(1983)}]{Pritchard:1983}%
  \BibitemOpen
  \bibfield  {author} {\bibinfo {author} {\bibfnamefont {D.~E.}\ \bibnamefont
  {Pritchard}},\ }\href@noop {} {\bibfield  {journal} {\bibinfo  {journal}
  {Physical Review Letters}\ }\textbf {\bibinfo {volume} {51}},\ \bibinfo
  {pages} {1336} (\bibinfo {year} {1983})}\BibitemShut {NoStop}%
\bibitem [{\citenamefont {Herschbach}\ \emph {et~al.}(2000)\citenamefont
  {Herschbach}, \citenamefont {Tol}, \citenamefont {Hogervorst},\ and\
  \citenamefont {Vassen}}]{Herschbach:2000}%
  \BibitemOpen
  \bibfield  {author} {\bibinfo {author} {\bibfnamefont {N.}~\bibnamefont
  {Herschbach}}, \bibinfo {author} {\bibfnamefont {P.~J.~J.}\ \bibnamefont
  {Tol}}, \bibinfo {author} {\bibfnamefont {W.}~\bibnamefont {Hogervorst}}, \
  and\ \bibinfo {author} {\bibfnamefont {W.}~\bibnamefont {Vassen}},\
  }\href@noop {} {\bibfield  {journal} {\bibinfo  {journal} {Physical Review
  A}\ }\textbf {\bibinfo {volume} {61}},\ \bibinfo {pages} {050702} (\bibinfo
  {year} {2000})}\BibitemShut {NoStop}%
\bibitem [{\citenamefont {Hijmans}\ \emph {et~al.}(1989)\citenamefont
  {Hijmans}, \citenamefont {Luiten}, \citenamefont {Setija},\ and\
  \citenamefont {Walraven}}]{Hijmans:1989}%
  \BibitemOpen
  \bibfield  {author} {\bibinfo {author} {\bibfnamefont {T.~W.}\ \bibnamefont
  {Hijmans}}, \bibinfo {author} {\bibfnamefont {O.~J.}\ \bibnamefont {Luiten}},
  \bibinfo {author} {\bibfnamefont {I.~D.}\ \bibnamefont {Setija}}, \ and\
  \bibinfo {author} {\bibfnamefont {J.~T.~M.}\ \bibnamefont {Walraven}},\
  }\href@noop {} {\bibfield  {journal} {\bibinfo  {journal} {JOSA B: Optical
  Physics}\ }\textbf {\bibinfo {volume} {6}},\ \bibinfo {pages} {2235}
  (\bibinfo {year} {1989})}\BibitemShut {NoStop}%
\bibitem [{\citenamefont {Schmidt}\ \emph {et~al.}(2003)\citenamefont
  {Schmidt}, \citenamefont {Hensler}, \citenamefont {Werner}, \citenamefont
  {Binhammer}, \citenamefont {G{\"o}rlitz},\ and\ \citenamefont
  {Pfau}}]{Schmidt:2003}%
  \BibitemOpen
  \bibfield  {author} {\bibinfo {author} {\bibfnamefont {P.~O.}\ \bibnamefont
  {Schmidt}}, \bibinfo {author} {\bibfnamefont {S.}~\bibnamefont {Hensler}},
  \bibinfo {author} {\bibfnamefont {J.}~\bibnamefont {Werner}}, \bibinfo
  {author} {\bibfnamefont {T.}~\bibnamefont {Binhammer}}, \bibinfo {author}
  {\bibfnamefont {A.}~\bibnamefont {G{\"o}rlitz}}, \ and\ \bibinfo {author}
  {\bibfnamefont {T.}~\bibnamefont {Pfau}},\ }\href@noop {} {\bibfield
  {journal} {\bibinfo  {journal} {JOSA B: Optical Physics}\ }\textbf {\bibinfo
  {volume} {20}},\ \bibinfo {pages} {960} (\bibinfo {year} {2003})}\BibitemShut
  {NoStop}%
\bibitem [{\citenamefont {Ketterle}\ and\ \citenamefont {van
  Druten}(1996)}]{Ketterle:1996}%
  \BibitemOpen
  \bibfield  {author} {\bibinfo {author} {\bibfnamefont {W.}~\bibnamefont
  {Ketterle}}\ and\ \bibinfo {author} {\bibfnamefont {N.~J.}\ \bibnamefont {van
  Druten}},\ }\href@noop {} {\bibfield  {journal} {\bibinfo  {journal}
  {Advances In Atomic, Molelcular, and Optical Physics}\ }\textbf {\bibinfo
  {volume} {37}},\ \bibinfo {pages} {181} (\bibinfo {year} {1996})}\BibitemShut
  {NoStop}%
\bibitem [{\citenamefont {Luiten}\ \emph {et~al.}(1996)\citenamefont {Luiten},
  \citenamefont {Reynolds},\ and\ \citenamefont {Walraven}}]{Luiten:1996}%
  \BibitemOpen
  \bibfield  {author} {\bibinfo {author} {\bibfnamefont {O.~J.}\ \bibnamefont
  {Luiten}}, \bibinfo {author} {\bibfnamefont {M.~W.}\ \bibnamefont
  {Reynolds}}, \ and\ \bibinfo {author} {\bibfnamefont {J.~T.~M.}\ \bibnamefont
  {Walraven}},\ }\href@noop {} {\bibfield  {journal} {\bibinfo  {journal}
  {Physical Review A}\ }\textbf {\bibinfo {volume} {53}},\ \bibinfo {pages}
  {381} (\bibinfo {year} {1996})}\BibitemShut {NoStop}%
\bibitem [{\citenamefont {Herschbach}\ \emph {et~al.}(2003)\citenamefont
  {Herschbach}, \citenamefont {Tol}, \citenamefont {Tychkov}, \citenamefont
  {Hogervorst},\ and\ \citenamefont {Vassen}}]{Herschbach:2003}%
  \BibitemOpen
  \bibfield  {author} {\bibinfo {author} {\bibfnamefont {N.}~\bibnamefont
  {Herschbach}}, \bibinfo {author} {\bibfnamefont {P.}~\bibnamefont {Tol}},
  \bibinfo {author} {\bibfnamefont {A.}~\bibnamefont {Tychkov}}, \bibinfo
  {author} {\bibfnamefont {W.}~\bibnamefont {Hogervorst}}, \ and\ \bibinfo
  {author} {\bibfnamefont {W.}~\bibnamefont {Vassen}},\ }\href@noop {}
  {\bibfield  {journal} {\bibinfo  {journal} {Jounal of Optics B: Quantum and
  Semiclassical Optics}\ }\textbf {\bibinfo {volume} {5}},\ \bibinfo {pages}
  {65} (\bibinfo {year} {2003})}\BibitemShut {NoStop}%
\bibitem [{\citenamefont {Herschbach}(2003)}]{Herschbach:2003:thesis}%
  \BibitemOpen
  \bibfield  {author} {\bibinfo {author} {\bibfnamefont {N.}~\bibnamefont
  {Herschbach}},\ }\emph {\bibinfo {title} {{Trapped Triplet Helium Atoms:
  Inelastic Collisions and Evaporative Cooling}}},\ \href@noop {} {Ph.D.
  thesis},\ \bibinfo  {school} {Vrije Universiteit Amsterdam} (\bibinfo {year}
  {2003})\BibitemShut {NoStop}%
\end{thebibliography}%

\end{document}